\documentclass[prd,twocolumn,showpacs]{revtex4-1}
\usepackage{amsmath}
\usepackage{latexsym}
\usepackage{amsfonts}
\usepackage{graphicx}
\usepackage{mathrsfs}
\usepackage{CJK}
\usepackage{longtable}

\usepackage{hyperref}
\hypersetup{
  colorlinks = true,
  urlcolor = blue,
  linkcolor = blue,
  citecolor = cyan,
  filecolor = magenta,
}

\newcommand{\script}{\mathscr}

\begin{document}
\begin{CJK*}{GB}{gbsn}

\title{Constraints on Horndeski Theory Using the Observations of Nordtvedt Effect, Shapiro Time Delay and Binary Pulsars}
\author{Shaoqi Hou}
\email{shou1397@hust.edu.cn}
\affiliation{School of Physics, Huazhong University of Science and Technology, Wuhan, Hubei 430074, China}
\author{Yungui Gong}
\email{yggong@hust.edu.cn}
\affiliation{School of Physics, Huazhong University of Science and Technology, Wuhan, Hubei 430074, China}
\date{\today}

\begin{abstract}
Alternative theories of gravity not only modify the polarization contents of the gravitational wave, but also affect the motions of the stars and the energy radiated away via the gravitational radiation.
These aspects leave imprints in the observational data, which enables the test of General Relativity and its alternatives.
In this work, the Nordtvedt effect and the Shapiro time delay  are  calculated in order to constrain Horndeski theory  using the observations of lunar laser ranging experiments and  Cassini time-delay data.
The effective stress-energy tensor  is also obtained using the method of Isaacson.
Gravitational wave radiation of a binary system is calculated, and the change of the period of a binary system is deduced for the elliptical orbit.
These results can be used to set constraints on Horndeski theory with the observations of binary systems, such as PSR J1738+0333.
Constraints have been obtained for some subclasses of Horndeski theory, in particular, those satisfying the gravitational wave speed limits from GW170817 and GRB 170817A.
\end{abstract}

\maketitle
\end{CJK*}

\section{Introduction}

General Relativity (GR) is one of the cornerstones of modern physics. However, it faces several challenges. For example, GR cannot be quantized, and it cannot explain the present accelerating expansion of universe, i.e., the problem of dark energy.
These challenges motivate the pursuit of the alternatives to GR, one of which is the scalar-tensor theory. The scalar-tensor theory contains a scalar field $\phi$ as well as a metric tensor $g_{\mu\nu}$ to describe the gravity. It is the simplest alternative metric theory of gravity.
It solves some of GR's problems. For example, the extra degree of freedom of the scalar field might account for the dark energy and explain the accelerating expansion of the universe.
Certain scalar-tensor theories can be viewed as the low energy limit of string theory, one of the candidates of quantum gravity \cite{Fujii:2003pa}.

The detection of gravitational waves by the Laser Interferometer Gravitational-Wave Observatory (LIGO) and Virgo confirms GR to an unprecedented precision \cite{Abbott:2016blz,Abbott:2016nmj,Abbott:2017vtc,Abbott:2017oio,TheLIGOScientific:2017qsa,Abbott:2017gyy} and also provides the possibility to test GR in the dynamical, strong field limit.
The recent GW170814  detected the polarizations for the first time, and the result showed that the pure tensor polarizations are favored against pure vector and pure scalar polarizations \cite{Abbott:2017oio}.
The newest GW170817 is the first neutron star-neutron star merger event, and the concomitant gamma-ray burst GRB 170817A was later observed by the {\it Fermi} Gamma-ray Burst Monitor and the Anti-Coincidence Shield for the Spectrometer for the International Gamma-Ray Astrophysics
Laboratory, independently \cite{TheLIGOScientific:2017qsa,Goldstein:2017mmi,Savchenko:2017ffs}.
This opens the new era of multi-messenger astrophysics.
It is thus interesting to study gravitational waves in alternative metric theories of gravity, especially the scalar-tensor theory.

In 1974, Horndeski \cite{Horndeski:1974wa} constructed the most general scalar-tensor theory whose action contains higher derivatives of $\phi$ and $g_{\mu\nu}$, but still yields at most the second order differential field equations, and thus has no Ostrogradsky instability \cite{Ostrogradsky:1850fid}.
Because of its generality, Horndeski theory includes several important specific theories, such as GR, Brans-Dicke theory \cite{Brans:1961sx},  and $f(R)$ gravity \cite{Buchdahl:1983zz,OHanlon:1972xqa,Teyssandier:1983zz} etc..

In Refs.~\cite{Liang:2017ahj,Hou:2017bqj,Gong:2017bru}, we discussed the gravitational wave solutions in $f(R)$ gravity and Horndeski theory, and their polarization contents.
These works showed that in addition to the familiar + and $\times$ polarizations in GR, there is a mixed state of the transverse breathing and longitudinal polarizations both excited by a massive scalar field, while a massless scalar field excites
the transverse breathing polarization only.
In this work, it will be shown that the presence of a dynamical scalar field also changes the amount of energy radiated away by the gravitational wave affecting,
for example, the inspiral of binary systems.
Gravitational radiation causes the damping of the energy of the binary system, leading to the change in the orbital period.
In fact, the first indirect evidence for the existence of gravitational waves is the decay of the orbital period of the Hulse-Taylor pulsar (PSR 1913+16) \cite{Hulse:1974eb}.

Previously, the effective stress energy tensor was obtained by Nutku \cite{1969ApJ...158..991N} using the method of Landau and Lifshitz \cite{LLvol2}.
The damping of a compact binary system due to gravitational radiation in Brans-Dicke theory was calculated in Refs.~\cite{Wagoner:1970vr,Will:1989sk,Damour:1998jk,Brunetti:1998cc}, then Alsing {\it et al.} \cite{Alsing:2011er} extended the analysis to the massive scalar-tensor theory.
Refs.~\cite{Stein:2010pn,Saffer:2017ywl} surveyed the effective stress-energy tensor for a wide class of alternative theories of gravity using several methods.
However, they did not consider Horndeski theory.
Refs.~\cite{Zhang:2017srh,Liu:2017xef} studied the gravitational radiation in screened modified gravity and $f(R)$ gravity.
Hohman \cite{Hohmann:2015kra} developed parameterized post-Newtonian (PPN) formalism for Horndeski theory.
In this work, the method of Isaacson is used to obtain the effective stress-energy tensor for Horndeski theory.
Then the effective stress-energy tensor is applied to calculate the rate of energy damping and the period change of a binary system, which can be compared with the observations on binary systems to constrain  Horndeski theory.
Nordtvedt effect and Shapiro time delay effect will also be considered to put further constraints.
Ashtekar and Bonga pointed out in Refs.~\cite{Ashtekar:2017ydh,Ashtekar:2017wgq} a subtle difference between the transverse-traceless part of $h_{\mu\nu}$ defined by $\partial^\nu h_{\mu\nu}=0,\,\eta^{\mu\nu}h_{\mu\nu}=0$ and the one defined by using the spatial transverse projector,
but this difference does not affect the energy flux calculated in this work.

There were constraints on  Horndeski theory and its subclasses in the past.
The observations of GW170817 and GRB 170817A put severe constraints on the speed of gravitational waves \cite{Monitor:2017mdv}.
Using this limit, Ref.~\cite{Creminelli:2017sry} required that $\partial G_5/\partial X$ $=0$ and $2\partial G_4/\partial X+\partial G_5/\partial \phi=0$, while
Ref.~\cite{Ezquiaga:2017ekz} required $\partial G_4/\partial X\approx0$ and $G_5\approx\text{constant}$.
Ref.~\cite{Baker:2017hug} obtained the similar results as Ref.~\cite{Ezquiaga:2017ekz}, and also pointed out that the self-accelerating theories should be shift symmetric.
Arai and Nishizawa  found that Horndeski theory with arbitrary functions $G_4$ and $G_5$ needs fine-tuning to account for the cosmic accelerating expansion \cite{Arai:2017hxj}.
For more constraints derived from the gravitational wave speed limit, please refer to Refs.~\cite{Sakstein:2017xjx,Gong:2017kim,Crisostomi:2017lbg}, and for more discussions on the constraints on the subclasses of Horndeski theory, please refer to Refs. \cite{Will:2014kxa,Lambiase:2015yia,Bhattacharya:2016naa,Banerjee:2017hzw,Shao:2017gwu}.

In this work, the calculation will be done in the Jordan frame, and the screening mechanisms, such as the chameleon \cite{Khoury:2003rn,Khoury:2003aq} and the symmetron \cite{Hinterbichler:2010es,Hinterbichler:2011ca}, are not considered.
Vainshtein mechanism was first discovered to solve the vDVZ discontinuity problem for massive gravity \cite{Vainshtein:1972sx}, and later found to also appear in  theories containing the derivative self-couplings of the scalar field, such as some subclasses of Horndeski theory \cite{Deffayet:2001uk,Babichev:2009ee,Koyama:2013paa,Babichev:2013usa,Winther:2015pta}.
When Vainshtein mechanism is in effect, the effect of nonlinearity cannot be ignored within the so-called Vainshtein radius $r_\text{V}$ from the center of the matter source.
Well beyond $r_\text{V}$, the linearization can be applied.
The radius $r_\text{V}$ depends on the parameters defining Horndeski theory, and can be much smaller than the size of a celestial object.
So in this work, we consider Horndeski theories which predict small $r_\text{V}$, if it exists, compared to the sizes of the Sun and neutron stars.
The linearization can thus be done even deep inside the stars.
In this case, one can safely ignore Vainshtein mechanism.

The paper is organized as follows.
In Section \ref{sechorneom}, Horndeski theory is briefly introduced and the equations of motion are derived up to the second order in perturbations around the flat spacetime background.
Section \ref{secest} derives the effective stress-energy tensor according to the procedure given by Isaacson.
Section \ref{secnew} is devoted to the computation of the metric and scalar perturbations in the near zone up to Newtonian order and the discussion of the motion of self-gravitating objects that source gravitational waves.
In particular, Nordtvedt effect and Shapiro time delay are discussed.
In Section \ref{secgws}, the metric and scalar perturbations are calculated in the far zone up to the quadratic order, and in Section \ref{secgwb}, these solutions are applied to a compact binary system to calculate the energy emission rate and the period change.
Section \ref{sec-exp} discusses the constraints on Horndeski theory based on the observations.
Finally, Section \ref{seccon} summarizes the results.
Throughout the paper, the speed of light in vacuum is taken to be $c=1$.

\section{Horndeski Theory}\label{sechorneom}

The  action of Horndeski theory is given by \cite{Kobayashi:2011nu},
\begin{equation}\label{acth}
  S=\int\mathrm{d}^4x\sqrt{-g}(\mathscr L_2+\mathscr L_3+\mathscr L_4+\mathscr L_5)+S_m[g_{\mu\nu},\psi_m],
\end{equation}
where $\psi_m$ represents matter fields, $S_m$ is the action for $\psi_m$, and the terms in the integrand are
\begin{gather}
  \mathscr L_2 = G_2(\phi,X),\,\script L_3=-G_3(\phi,X)\Box\phi, \\
  \script L_4 = G_4(\phi,X)R+G_{4X}[(\Box\phi)^2-(\phi_{;\mu\nu})^2], \\
  \script L_5 = G_5(\phi,X)G_{\mu\nu}\phi^{;\mu\nu}-\frac{G_{5X}}{6}[(\Box\phi)^3-3(\Box\phi)(\phi_{;\mu\nu})^2\nonumber\\
  +2(\phi_{;\mu\nu})^3].
\end{gather}
In these expressions, $X=-\phi_{;\mu}\phi^{;\mu}/2$ with $\phi_{;\mu}=\nabla_\mu\phi$, $\phi_{;\mu\nu}=\nabla_\nu\nabla_\mu\phi$, $\Box\phi=g^{\mu\nu}\phi_{;\mu\nu}$, $(\phi_{;\mu\nu})^2=\phi_{;\mu\nu}\phi^{;\mu\nu}$ and $(\phi_{;\mu\nu})^3=\phi_{;\mu\nu}\phi^{;\mu\rho}\phi^{;\nu}{}_{;\rho}$ for simplicity. $G_i\, (i=2,3,4,5)$ are arbitrary functions of $\phi$ and $X$ \footnote{$G_2$ is usually called $K_2$ in literature.}.
For notational simplicity and clarity, we define the following symbol for the function $f(\phi,X)$,
\begin{equation}\label{defsym}
  f_{(m,n)}=\frac{\partial^{m+n}f(\phi,X)}{\partial \phi^m\partial X^n}\Big|_{\phi=\phi_0,X=0},
\end{equation}
so in particular, $f_{(0,0)}=f(\phi_0,0)$ with $\phi_0$ the value of $\phi$ in the flat spacetime background.

Suitable choices of $G_i$ reproduce interesting subclasses of Horndeski theory.
For instance, one obtains GR by choosing $G_4=(16\pi G_\mathrm{N})^{-1}$ and the remaining $G_i=0$, with $G_\mathrm{N}$ Newton's constant.
Brans-Dicke theory is recovered with $G_2=2\omega_\text{BD} X/\phi, G_4=\phi, G_3=G_5=0$, while the massive scalar-tensor theory with a potential $U(\phi)$ \cite{Alsing:2011er}
is obtained with $G_2=2\omega_\text{BD} X/\phi-U(\phi),\,G_4=\phi,\,G_3=G_5=0$, where $\omega_\text{BD}$ is a constant; or with
$G_2=X-U(\phi)$, $G_4=g(\phi)$, $G_3=G_5=0$.
Finally, $f(R)$ gravity is given by $G_2=f(\phi)-\phi f'(\phi)$, $G_4=f'(\phi)$, $G_3=G_5=0$ with $f'(\phi)=\mathrm{d} f(\phi)/\mathrm{d}\phi$.

\subsection{Matter action}

Although there are no coupling terms between matter fields $\psi_m$ and $\phi$, matter fields $\psi_m$ indirectly interact with $\phi$ via the metric tensor.
For example, in Brans-Dicke theory, $\phi$ acts effectively like the gravitational constant, which influences the internal structure and motion of a gravitating object, so the binding energy of the object depends on $\phi$.
Since the total energy $E$ is related to the inertial mass $m$, then $m$ depends on $\phi$, too.
When their spins and multipole moments can be ignored, the gravitating objects can be described by point like particles, and the effect of $\phi$ can be taken into account by the following matter action according to Eardley's prescription \cite{1975ApJ...196L..59E},
\begin{equation}\label{matlag}
  S_m=-\sum_a\int m_a(\phi)\mathrm{d}\tau_a,
\end{equation}
whose stress-energy tensor is
\begin{equation}\label{matset}
  T_{\mu\nu}=\frac{1}{\sqrt{-g}}\sum_am_a(\phi)\frac{u_\mu u_\nu}{u^0}\delta^{(4)}(x^\lambda-x^\lambda_a(\tau)),
\end{equation}
where $x^\lambda_a(\tau)$ describes the worldline of particle $a$ and $u^\mu=\mathrm{d} x^\mu(\tau)/\mathrm{d}\tau$.
Therefore, if there is no force other than  gravity acting on a self-gravitating object, this object will not follow the geodesic. This causes the violation of the strong equivalence principle (SEP).

In this work, the gravitational wave is studied in the flat spacetime background with $g_{\mu\nu}=\eta_{\mu\nu}$ and $\phi=\phi_0$,
so we expand the masses around the value $\phi_0$ in the following way,
\begin{equation}\label{expmasses}
  m_a(\phi)=m_a\left[1+\frac{\varphi}{\phi_0}s_a-\frac{1}{2}\left(\frac{\varphi}{\phi_0}\right)^2(s_a'-s_a^2+s_a)+O(\varphi^3)\right].
\end{equation}
Here, $\varphi=\phi-\phi_0$ is the perturbation, and $m_{a}=m_a(\phi_0)$ for simplicity.
This expansion also requires that $\phi_0\ne0$, so the present discussion does not apply to $f(R)$ gravity.
$s_a$ and $s'_a$ are the first and second sensitivities of the mass $m_a$,
\begin{equation}\label{defss}
  s_a=\frac{\mathrm{d}\ln m_a(\phi)}{\mathrm{d}\ln\phi}\Big|_{\phi_0},\quad s'_a=-\frac{\mathrm{d}^2\ln m_a(\phi)}{\mathrm{d}(\ln\phi)^2}\Big|_{\phi_0}.
\end{equation}
The sensitivities measure the violation of SEP.

\subsection{Linearized equations of motion}

The equations of motion can be obtained and simplified using {\it xAct} package \cite{Martin-Garcia:2007bqa,MartinGarcia:2008qz,xperm2008,Brizuela:2008ra,xact}.
Because of their tremendous complexity, the full equations of motion will not be presented.
Interested readers are referred to Refs.~\cite{Kobayashi:2011nu,Gao:2011mz}.
As we checked, {\it xAct} package gives the same equations of motion as Refs.~\cite{Kobayashi:2011nu,Gao:2011mz}.
For the purpose of this work, the equations of motion are expanded up to the second order in perturbations defined as
\begin{equation}\label{defpert}
  g_{\mu\nu}=\eta_{\mu\nu}+h_{\mu\nu},\quad \phi=\phi_0+\varphi.
\end{equation}
These equations are given in \ref{app-eoms2nd}.

The gravitational wave solutions are investigated in the flat spacetime background, which requires that
\begin{equation}\label{flatre}
  G_{2(0,0)}=0,\quad G_{2(1,0)}=0.
\end{equation}
This can be easily checked by a quick inspection of Eqs. \eqref{eq-eineomupto2} and \eqref{eq-phieomupto2}.
Then dropping higher order terms in Eqs. \eqref{eq-eineomupto2} and \eqref{eq-phieomupto2}, the linearized equations of motion are thus given by
\begin{gather}
  (G_{2(0,1)}-2G_{3(1,0)})\Box\varphi+G_{2(2,0)}\varphi +G_{4(1,0)}R^{(1)}= -\left(\frac{\partial T}{\partial \phi}\right)^{(1)},\label{eq-schgb} \\
  G_{4(0,0)}G_{\mu\nu}^{(1)}-G_{4(1,0)}(\partial_\mu\partial_\nu\varphi-\eta_{\mu\nu}\Box\varphi) = \frac{1}{2}T_{\mu\nu}^{(1)},\label{eq-einhgb}
\end{gather}
where $T=g^{\mu\nu}T_{\mu\nu}$ is the trace, $\Box=\eta^{\mu\nu}\partial_\mu\partial_\nu$ from now on, and the superscript $(1)$ implies the leading order part of the quantity.

The equations of motion can be decoupled by introducing an auxiliary field $\tilde h_{\mu\nu}$  defined  as following,
\begin{equation}\label{auht}
  \tilde h_{\mu\nu}=h_{\mu\nu}-\frac{1}{2}\eta_{\mu\nu}h-\frac{G_{4(1,0)}}{G_{4(0,0)}}\eta_{\mu\nu}\varphi,
\end{equation}
where $h=\eta^{\mu\nu} h_{\mu\nu}$ is the trace,
and the original metric tensor perturbation is,
\begin{equation}\label{hinht}
 h_{\mu\nu}=\tilde h_{\mu\nu}-\frac{1}{2}\eta_{\mu\nu}\tilde h-\frac{G_{4(1,0)}}{G_{4(0,0)}}\eta_{\mu\nu}\varphi,
\end{equation}
with $\tilde h=\eta^{\mu\nu}\tilde h_{\mu\nu}$.
The equations of motion are gauge invariant under the the following infinitesimal coordinate transformation,
\begin{equation}\label{gatrans}
  \varphi' = \varphi,\quad\tilde h'_{\mu\nu} = \tilde h_{\mu\nu}-\partial_\mu\xi_\nu-\partial_\nu\xi_\mu+\eta_{\mu\nu}\partial_\rho\xi^\rho,
\end{equation}
with $x'^\mu=x^\mu+\xi^\mu$.
Therefore, one can choose the transverse gauge $\partial_\nu \tilde h^{\mu\nu}=0$, and after some algebraic manipulations, the equations of motion become
\begin{gather}
  (\Box-m^2_s)\varphi = \frac{T^{(1)}_*}{2G_{4(0,0)}\zeta}, \label{eq-sceqf}\\
  \Box\tilde h_{\mu\nu} = -\frac{T^{(1)}_{\mu\nu}}{G_{4(0,0)}},\label{eq-eineqf}
\end{gather}
where  $T_*^{(1)}=G_{4(1,0)}T^{(1)}-2G_{4(0,0)}(\partial T/\partial \phi)^{(1)}$ \footnote{The way defining $T_*^{(1)}$ is different from the one defining $T^*$ in Ref.~\cite{Alsing:2011er} in that the coefficient of $T^{(1)}_*$ is not 1.}
with $T^{(1)}=\eta^{\mu\nu}T_{\mu\nu}^{(1)}$, and the mass of the scalar field is
\begin{gather}\label{msq}
  m^2_s=-G_{2(2,0)}/\zeta,\nonumber\\
   \zeta=G_{2(0,1)}-2G_{3(1,0)}+3G_{4(1,0)}^2/G_{4(0,0)}.
\end{gather}
Of course, $\zeta\ne0$, otherwise $\varphi$ is non-dynamical.

From the equations of motion \eqref{eq-sceqf} and \eqref{eq-eineqf}),
one concludes that the scalar field is generally massive unless $G_{2(2,0)}$ is zero,
and the auxiliary field $\tilde h_{\mu\nu}$ resembles the spin-2 graviton field $\bar h_{\mu\nu}=h_{\mu\nu}-\eta_{\mu\nu}h/2$ in GR.
$\tilde h_{\mu\nu}$ is sourced by the matter stress-energy tensor, while the source of the scalar perturbation $\varphi$ is a linear combination of the trace of the matter stress-energy tensor and the partial derivative of the trace with respect to $\phi$.
This is because of the indirect interaction between the scalar field and the matter field via the metric tensor.

\section{Effective Stress-Energy Tensor}\label{secest}

The method of Isaacson \cite{Isaacson:1967zz,Isaacson:1968zza} will be used to obtain the effective stress-energy tensor for gravitational waves in Horndeski theory in the short-wavelength approximation, i.e., the wavelength $\lambda\ll 1/\sqrt{R}$ with $R$ representing the typical value of the background Riemann tensor components.
This approximation is trivially satisfied in our case, as the background is flat and $R=0$.
In averaging over several wavelengths, the following rules are utilized \cite{mtw}:
\begin{enumerate}
  \item The average of a gradient is zero, e.g., $\langle\partial_\mu(\tilde h_{\rho\sigma}\partial_\nu\tilde h)\rangle=0$,
  \item One can integrate by parts, e.g., $\langle\tilde h\partial_{\rho}\partial_{\sigma}\tilde h_{\mu\nu}\rangle=$ $-\langle\partial_\rho\tilde h$ $\partial_\sigma\tilde h_{\mu\nu}\rangle$,
\end{enumerate}
where $\langle\,\rangle$ implies averaging. These rules apply to not only terms involving $\tilde h$ but also those involving $\varphi$.
In the case of a curved background, these rules are supplemented by the one that covariant derivatives commute, which always holds in the flat background case.

With this method, the effective stress-energy tensor in an arbitrary gauge can be calculated straightforwardly using {\it xAct} and given by,
\begin{equation}\label{effset}
\begin{split}
 & T_{\mu\nu}^\mathrm{GW}=\\
 &\left\langle\frac{1}{2}G_{4(0,0)}\Big(\partial_\mu\tilde h_{\rho\sigma}\partial_\nu\tilde h^{\rho\sigma}-\frac{1}{2}\partial_\mu\tilde h\partial_\nu\tilde h-\partial_\mu\tilde h_{\nu\rho}\partial_\sigma\tilde h^{\sigma\rho}\right.\\
  &-\partial_\nu\tilde h_{\mu\rho}\partial_\sigma\tilde h^{\sigma\rho}\Big)\\
  &+\zeta\partial_\mu\varphi\partial_\nu\varphi\\
  &\left.+G_{4(1,0)}(m_s^2\varphi\tilde h_{\mu\nu}+\partial_\mu\varphi\partial^\rho\tilde h_{\rho\nu}+\partial_\nu\varphi\partial^\rho\tilde h_{\rho\mu}\right.\\
  &\left.-\eta_{\mu\nu}\partial_\sigma\varphi\partial_\rho\tilde h^{\rho\sigma})\right\rangle.
\end{split}
\end{equation}
It can be checked that this expression is gauge invariant under Eq.~(\ref{gatrans}).
In fact, the terms in the first around brackets take exactly the same forms as in GR excerpt for a different factor.
The fourth line remains invariant, as $\varphi'=\varphi$ in the gauge transformation.
To show that the remaining lines are also gauge invariant, making the replacement $\tilde h_{\mu\nu}\rightarrow\tilde h_{\mu\nu}-\partial_\mu\xi_\nu-\partial_\nu\xi_\mu+\eta_{\mu\nu}\partial_\rho\xi^\rho$ gives
\begin{equation}\label{eq-ch-ginv}
\begin{split}
  &\text{Remaining lines}=\\
  &\Big\langle G_{4(1,0)}(m_s^2\varphi\tilde h_{\mu\nu}+\partial_\mu\varphi\partial^\rho\tilde h_{\rho\nu}+\partial_\nu\varphi\partial^\rho\tilde h_{\rho\mu}\\
  &-\eta_{\mu\nu}\partial_\sigma\varphi\partial_\rho\tilde h^{\rho\sigma})\Big\rangle\\
  &+\Big\langle m_s^2G_{4(1,0)}\varphi(-\partial_\mu\xi_\nu-\partial_\nu\xi_\mu+\eta_{\mu\nu}\partial_\rho\xi^\rho)\\
  &+G_{4(1,0)}(-\partial_\mu\varphi\partial_\rho\partial^\rho\xi_\nu-\partial_\nu\varphi\partial_\rho\partial^\rho\xi_\mu\\
  &+\eta_{\mu\nu}\partial_\sigma\varphi\partial_\rho\partial^\rho\xi^\sigma)\Big\rangle.
  \end{split}
\end{equation}
Far away from the matter, $\partial_\rho\partial^\rho\varphi=m_s^2\varphi$ according to Eq.~\eqref{eq-sceqf}.
Substituting this into the fourth line of Eq.~\eqref{eq-ch-ginv}, one immediately finds total derivatives of the forms $\partial_\mu(\varphi\partial_\rho\partial^\rho\xi_\nu)$ and $\partial_\sigma(\varphi\partial_\rho\partial^\rho\xi^\sigma)$.
So the first averaging rule  implies that the last three lines of Eq.~\eqref{eq-ch-ginv} vanish.
Therefore, the effective stress-energy tensor \eqref{effset} is indeed gauge invariant.

In vacuum, the transverse-traceless (TT) gauge ($\partial_\nu \tilde h^{\mu\nu}=0$ and $\tilde h=0$) can be taken, and the effective stress-energy simplifies,
\begin{equation}\label{effsettt}
\begin{split}
  T_{\mu\nu}^\mathrm{GW}=&\left\langle\frac{1}{2}G_{4(0,0)}\partial_\mu\tilde h_{\rho\sigma}^\mathrm{TT}\partial_\nu\tilde h^{\rho\sigma}_\mathrm{TT}+\zeta\partial_\mu\varphi\partial_\nu\varphi\right.\\
  &\left.+m_s^2G_{4(1,0)}\varphi\tilde h_{\mu\nu}^\mathrm{TT}\right\rangle,
\end{split}
\end{equation}
where $\tilde h_{\mu\nu}^\mathrm{TT}$ denotes the transverse-traceless part.
In the limit that $G_4=(16\pi G_\mathrm{N})^{-1}$ and the remaining arbitrary functions $G_i$ vanish, Eq.~(\ref{effset}) recovers the effective stress-energy tensor of GR \cite{mtw}.
One can also check that Eq.~(\ref{effset}) reduces to the one given in Ref.~\cite{Brunetti:1998cc} for Brans-Dicke theory in the gauge of $\partial_\nu\tilde h^{\mu\nu}=0$ and $\tilde h=-2\varphi/\phi_0$.

In order to calculate the energy carried away by gravitational waves, one has to first study the motion of the source. This is the topic of the next section.

\section{The Motion of Gravitating Objects in the Newtonian Limit}\label{secnew}

The motion of the source will be calculated in the Newtonian limit.
The source is modeled as a collection of gravitating objects with the action given by Eq.~(\ref{matlag}). In the slow motion, weak field limit, there exists a nearly global inertial reference frame. In this frame, a Cartesian
coordinate system is established whose origin is chosen to be the center of mass of the matter source. Let $\vec x$ represent the field point whose length is denoted by $r=|\vec x|$.

In the near zone \cite{Poisson2014}, the metric and the scalar perturbations will be calculated at the Newtonian order.
The stress-energy tensor of the matter source is given by \footnote{The matter stress-energy tensor $T_{\mu\nu}$ and the derivative of its trace $T$ with respect to $\phi$, $\partial T/\partial\phi$, are both expanded beyond the leading order, because the higher order contributions are need to calculate the scalar perturbations in Section \ref{secgws}.},
\begin{equation}\label{setapprox}
\begin{split}
  T_{\mu\nu}=&\sum_am_{a}u_\mu u_\nu \left(1-\frac{1}{2} v_a^2\right.\\
  &\left.-\frac{1}{2}h_{jj}+s_a\frac{\varphi}{\phi_0}+O(v^4)\right)\delta^{(4)}(x^\lambda-x^\lambda_a(\tau)),
\end{split}
\end{equation}
and one obtains,
\begin{equation}\label{dsetapprox}
\begin{split}
  \frac{\partial T}{\partial \phi}=&-\sum_a\frac{m_a}{\phi_0}\left[s_a\left(1-\frac{1}{2}h_{jj}-\frac{v_a^2}{2}\right)\right.\\
  &\left.-(s_a'-s_a^2+s_a)\frac{\varphi}{\phi_0}+O(v^4)\right]\delta^{(4)}(x^\lambda-x^\lambda_a(\tau)).
\end{split}
\end{equation}
In these expressions, the 4-velocity of particle $a$ is $u_a^\mu=u_a^0(1,\vec v_a)$ and $v_a^2=\vec v_a^2$.
With these results, the leading order of the source for the scalar field is
\begin{equation}\label{srcscl}
\begin{split}
  T_*^{(1)}=&-\sum_am_{a}S_a\delta^{(4)}(x^\lambda-x^\lambda_a(\tau)),
\end{split}
\end{equation}
with  $S_a=G_{4(1,0)}-\frac{2G_{4(0,0)}}{\phi_0}s_a$.

Now, the linearized equations (\ref{eq-sceqf}, \ref{eq-eineqf}) take the following forms
\begin{gather}
   (\Box-m_s^2)\varphi =- \frac{1}{2G_{4(0,0)}\zeta}\sum_am_{a}S_a\delta^{(4)}(x^\lambda-x^\lambda_a(\tau)), \label{eq-sceqfex}\\
  \Box\tilde h_{\mu\nu} = -\frac{1}{G_{4(0,0)}}\sum_am_{a}u_\mu u_\nu\delta^{(4)}(x^\lambda-x^\lambda_a(\tau)),\label{eq-eineqfex}
\end{gather}
and the leading order contributions to the perturbations are easily obtained,
\begin{gather}
  \varphi(t,\vec x)= \frac{1}{8\pi G_{4(0,0)}\zeta}\sum_a\frac{m_aS_a}{r_a}e^{-m_sr_a},\label{solscllo}\\
  \tilde h_{00}(t,\vec x)=\frac{1}{4\pi G_{4(0,0)}}\sum_a\frac{m_a}{r_a},
\end{gather}
and $\tilde h_{0j}=\tilde h_{jk}=0$ at this order, where $r_a=|\vec x-\vec x_a|$ and the scalar field is given by a sum of Yukawa potentials.
The leading order metric perturbation can be determined by  Eq.~(\ref{hinht}),
\begin{gather}
  h_{00}=\frac{1}{8\pi G_{4(0,0)}}\sum_a\frac{m_a}{r_a}\left(1+\frac{G_{4(1,0)}}{G_{4(0,0)}\zeta}S_ae^{-m_sr_a}\right), \label{solh00lo}\\
  h_{jk}=\frac{\delta_{jk}}{8\pi G_{4(0,0)}}\sum_a\frac{m_a}{r_a}\left(1-\frac{G_{4(1,0)}}{G_{4(0,0)}\zeta}S_ae^{-m_sr_a}\right)\label{solhjklo},
\end{gather}
with $h_{0j}=0$.

\subsection{Static, spherically symmetric solutions}

For the static, spherically symmetric solution with a single point mass $M$ at rest at the origin as the source,
the time-time component of the metric tensor is
\begin{equation}\label{ttmet}
  g_{00}=-1+\frac{1}{8\pi G_{4(0,0)}}\frac{M}{r}\left(1+\frac{G_{4(1,0)}}{G_{4(0,0)}\zeta}S_Me^{-m_sr}\right)+\cdots,
\end{equation}
where $S_M=G_{4(1,0)}-2G_{4(0,0)}s_M/\phi_0$ and $s_M$ is the sensitivity of the point mass $M$. From this, the ``Newton's constant" can be read off
\begin{equation}\label{newc}
  G_\mathrm{N}(r)=\frac{1}{16\pi G_{4(0,0)}}\left(1+\frac{G_{4(1,0)}}{G_{4(0,0)}\zeta}S_Me^{-m_sr}\right),
\end{equation}
which actually depends on the distance $r$ because the scalar field is massive.
The measured Newtonian constant at the earth is $G_\mathrm{N}(r_\otimes)$ with $r_\otimes$ the radius of the Earth.
The ``post-Newtonian parameter" $\gamma(r)$ can also be read off by examining $g_{jk}$, which is
\begin{equation}\label{jkmet}
\begin{split}
  g_{jk}=&\delta_{jk}\left[1+\frac{1}{8\pi G_{4(0,0)}}\frac{M}{r}\left(1-\frac{G_{4(1,0)}}{G_{4(0,0)}\zeta}S_Me^{-m_sr}\right)\right]+\cdots\\
  =&\delta_{jk}\left(1+2\frac{G_{4(0,0)}\zeta-G_{4(1,0)}S_Me^{-m_sr}}{G_{4(0,0)}\zeta+G_{4(1,0)}S_Me^{-m_sr}}G_\mathrm{N}(r)\frac{M}{r}\right)+\cdots.
  \end{split}
\end{equation}
In the PPN formalism, the space-space components of the metric take the following form,
\begin{equation}\label{ppnmet}
  g_{jk}^\mathrm{PPN}=\delta_{jk}\left(1+2\gamma G_\mathrm{N}\frac{M}{r}\right)+\cdots,
\end{equation}
where the parameter $\gamma$ is a constant. So
\begin{equation}\label{ppngamma}
  \gamma(r)=\frac{G_{4(0,0)}\zeta-G_{4(1,0)}S_Me^{-m_sr}}{G_{4(0,0)}\zeta+G_{4(1,0)}S_Me^{-m_sr}}.
\end{equation}
The above result can recover the results for $f(R)$ gravity and general scalar-tensor theory \cite{Capone:2009xk,Perivolaropoulos:2009ak,Hohmann:2013rba,Hohmann:2015kra}
if we keep the equivalence principle.
In the massless case ($G_{2(2,0)}=0$), we get
\begin{gather}
  G_\mathrm{N}= \frac{1}{16\pi G_{4(0,0)}}\left[1+\frac{G_{4(1,0)}}{G_{4(0,0)}\zeta}S_M\right],\label{eq-gn-ml}\\
  \gamma=\frac{G_{4(0,0)}\zeta-G_{4(1,0)}S_M}{G_{4(0,0)}\zeta+G_{4(1,0)}S_M}.\label{eq-gam-ml}
\end{gather}
Note that $G_\mathrm{N}(r)$ and $\gamma(r)$ both depend on $S_M$ which reflects the internal structure and motion of the gravitating object in question.
Even if the scalar field is massless, this dependence still persists.
Therefore, neither of them is universal due to the violation of SEP caused by the scalar field.
It is obvious that $G_\text{N}(r_\otimes)$ should take the same value as $G_\mathrm{N}$.

\subsection{Equations of motion of the matter}

With the near zone solutions \eqref{solscllo}, \eqref{solh00lo} and \eqref{solhjklo}
one obtains the total matter Lagrangian up to the linear order,
\begin{equation}\label{matlaglo}
\begin{split}
  L_\mathrm{m}=&-\sum_am_a\Bigg[1-\frac{1}{2}v_a^2\\
  &-\frac{1}{32\pi G_{4(0,0)}}\sum_{b\ne a}\frac{m_b}{r_{ab}}\left(1+\frac{S_aS_b}{G_{4(0,0)}\zeta} e^{-m_sr_{ab}}\right)\Bigg],
\end{split}
\end{equation}
where $r_{ab}=|\vec{x}_a-\vec x_b|$ is the distance between the particles $a$ and $b$. The equation of motion for the mass $m_a$ can thus be obtained using the Euler-Lagrange equation, yielding its acceleration,
\begin{equation}\label{acca}
\begin{split}
  a_a^j=&-\frac{1}{16\pi G_{4(0,0)}}\sum_{b\ne a}\frac{m_b}{r_{ab}^2}\hat r^j_{ab}\times\\
  &\left[1+\frac{S_aS_b}{G_{4(0,0)}\zeta}(1+m_sr_{ab})e^{-m_sr_{ab}}\right],
\end{split}
\end{equation}
with $\hat r_{ab}=(\vec x_a-\vec x_b)/r_{ab}$. In particular, for a binary system, the relative acceleration $a^j=a^j_1-a^j_2$ is
\begin{equation}\label{relaccb}
  a^j=-\frac{m\hat r^j_{12}}{16\pi G_{4(0,0)}r_{12}^2}\left[1+\frac{S_aS_b}{G_{4(0,0)}\zeta}(1+m_sr_{12})e^{-m_sr_{12}}\right],
\end{equation}
where $m=m_1+m_2$ is the total mass. The first term in the square brackets gives the result that resembles the familiar Newtonian gravitational acceleration,
while the second one reflects the effect of the scalar field.
In the massless case, the second term no longer depends on $r_{12}$ and can be absorbed into the first one, so the binary system moves in a similar way as in Newtonian gravity with a modified Newton's constant.

The Hamiltonian of the matter is
\begin{equation}\label{mham}
\begin{split}
  H_\mathrm{m}=&\sum_a\vec p_a\cdot\vec x_a-L_\text{m}\\
  =&\sum_am_a\Bigg[\frac{1}{2}v_a^2-\frac{1}{32\pi G_{4(0,0)}}\times\\
  &\sum_{b\ne a}\frac{m_b}{r_{ab}}\left(1+\frac{S_aS_b}{G_{4(0,0)}\zeta}e^{-m_sr_{ab}}\right)\Bigg],
  \end{split}
\end{equation}
where $p_a^j=\partial L_\text{m}/\partial x_a^j$ is the $j$-th component of the canonical momentum of particle $a$, and the total rest mass has been dropped.
In particular, the Hamiltonian of a binary system is given by
\begin{equation}\label{mhamb}
\begin{split}
  H_\mathrm{m}=&\frac{\mu v^2}{2}-\frac{\mu m}{16\pi G_{4(0,0)}r_{12}}\times\\
  &\left[1+
  \frac{S_1S_2}{G_{4(0,0)}\zeta}(1+m_sr_{12})e^{-m_sr_{12}}\right],
\end{split}
\end{equation}
where $\vec v=\vec v_1-\vec v_2$, and $\mu=m_1m_2/m$ is the reduced mass.
This will be useful for calculating the total mechanical energy of a binary system and the ratio of energy loss due to the gravitational radiation.

\subsection{Nordtvedt effect}

The presence of the scalar field modifies the trajectories of self-gravitating bodies.
They will no longer follow  geodesics.
Therefore, SEP is violated in Horndeski theory.
This effect is called the Nordtvedt effect \cite{Nordtvedt:1968qr,Nordtvedt:1968qs}. It results in measurable effects in the solar system, one of which is the polarization of the Moon's orbit around the Earth \cite{1982RPPh...45..631N,Will:1993ns}.

To study the Nordtvedt effect, one considers a system of three self-gravitating objects $a,\,b$ and $c$ and studies the relative acceleration of $a$ and $b$ in the field of $c$.
With Eq.~(\ref{acca}) and assuming $r_{ab}\ll r_{ac}\approx r_{bc}$, the relative acceleration is
\begin{equation}\label{relaccabapp}
\begin{split}
  a_{ab}^j\approx&-\frac{1}{16\pi G_{4(0,0)}}\frac{m_a+m_b}{r_{ab}^2}\hat r^j_{ab}\times\\
  &\left[1+\frac{S_aS_b}{G_{4(0,0)}\zeta}(1+m_sr_{ab})e^{-m_sr_{ab}}\right]\\
  &-\frac{m_c}{16\pi G_{4(0,0)}}\left(\frac{\hat r^j_{ac}}{r_{ac}^2}-\frac{\hat r^j_{bc}}{r_{bc}^2}\right)\\
  &+\frac{S_c(s_a-s_b)}{8\pi G_{4(0,0)}\phi_0\zeta}\frac{m_c\hat r_{ac}^j}{r_{ac}^2}(1+m_sr_{ac})e^{-m_sr_{ac}},
\end{split}
\end{equation}
where the first term presents the Newtonian acceleration modified by the presence of the scalar field, the second is the tidal force caused by the gravitational gradient due to the object $c$,
and the last one describes the Nordtvedt effect. The effective Nordtvedt parameter is
\begin{equation}\label{effNordpar}
  \eta_\mathrm{N}=\frac{S_c}{8\pi G_\text{N}G_{4(0,0)}\phi_0\zeta}(1+m_sr_{ac})e^{-m_sr_{ac}}.
\end{equation}
This parameter depends on $S_c=G_{4(1,0)}-2G_{4(0,0)}s_c/\phi_0$, so this effect is indeed caused by the violation of SEP.

\subsection{Shapiro time delay effect}

Another effect useful for constraining Horndeski theory is the Shapiro time delay \cite{Shapiro:1964uw}. In order to calculate this effect, one considers the photon propagation time in a static (or nearly static)
gravitational field produced by a single mass $M$ at the origin. Due to the presence of gravitational potential, the 3-velocity of the photon in the nearly inertial coordinate
system is no longer 1 and varies. The propagation time is thus different from that when the spacetime is flat.
Let the 4 velocity of the photon be $u^\mu=u^0(1,\vec v)$, then $u^\mu u_\mu=0$ gives
\begin{equation}\label{normph}
  -1+h_{00}+(\delta_{jk}+h_{jk})v^j v^k=0,
\end{equation}
where $h_{00}$ and $h_{jk}$ are given by Eqs.~\eqref{solh00lo} and \eqref{solhjklo} specialized to a single mass $M$ case.
In the flat spacetime, the trajectory for a photon emitted from position $\vec x_e$ at time $t_e$ is a straight line $\vec x(t)=\vec x_e+\hat N(t-t_e)$, where
$\hat N$ is the direction of the photon.
The presence of the gravitational potential introduces a small perturbation $\delta\vec x(t)$ so that $\vec x(t)=\vec x_e+\hat N(t-t_e)+\delta\vec x(t)$.
Substituting Eqs.~\eqref{solh00lo} and \eqref{solhjklo} into Eq.~(\ref{normph}), one obtains
\begin{equation}\label{vardelx}
  \hat N\cdot\frac{\mathrm{d}\delta \vec x}{\mathrm{d} t}=-\frac{M}{8\pi G_{4(0,0)}r(t)},
\end{equation}
where $r(t)=|\vec x(t)|$.
Suppose the photon emitted from position $\vec x_e$ is bounced back at position $\vec x_p$ and finally returns to $\vec x_e$. The total propagation time is
\begin{equation}\label{totpt}
  \Delta t=2|\vec x_p-\vec x_e|+\delta t,
\end{equation}
where $\delta t$ is caused by the Shapiro time delay effect,
\begin{equation}\label{shatimdel}
\begin{split}
  \delta t=&2\int_{t_e}^{t_p}\hat N\cdot\frac{\mathrm{d}\delta \vec x}{\mathrm{d} t}\mathrm{d} t\\
  =&\frac{M}{4\pi G_{4(0,0)}}\ln\frac{(r_e+\hat N\cdot\vec x_e)(r_p-\hat N\cdot\vec x_p)}{r_b^2},
  \end{split}
\end{equation}
where $r_e=|\vec x_e|$, $r_p=|\vec x_p|$ and $r_b=|\hat N\times \vec x_e|$ is the impact parameter of the photon relative to the source.

Since $M$ in Eq.~(\ref{shatimdel}) is not measurable, one replaces it with the Keplerian mass
\begin{equation}\label{defKm}
  M_\mathrm{K}=\frac{M}{16\pi G_{4(0,0)}G_\mathrm{N}}\left(1+\frac{G_{4(1,0)}}{G_{4(0,0)}\zeta}S_Me^{-m_sr}\right),
\end{equation}
with $S_M=G_{4(1,0)}-2G_{4(0,0)}s_M/\phi_0$ and $s_M$ the sensitivity of the source. In terms of $M_\mathrm{K}$, the Shapiro time delay is
\begin{equation}\label{shatimdel2}
  \delta t=2G_\mathrm{N}M_\mathrm{K}(1+\gamma(r))\ln\frac{(r_e+\hat N\cdot\vec x_e)(r_p-\hat N\cdot\vec x_p)}{r_b^2}.
\end{equation}
For the Shapiro time delay occurring near the Sun, $r$ in the above equation should be 1 AU, as this is approximately the distance where the Keplerian mass $M_\text{K}$ of the Sun is measured.

\section{Gravitational Wave Solutions}\label{secgws}

In the far zone, only the space-space components of the metric perturbation are needed to calculate the effective stress-energy tensor.
Since  the equation of motion (\ref{eq-eineqf}) for $\tilde h_{\mu\nu}$  takes the similar form as in GR, the leading order contribution to $\tilde h_{jk}$ is given by,
\begin{equation}\label{hjklo}
  \tilde h_{jk}(t,\vec x)=\frac{1}{8\pi G_{4(0,0)}r}\frac{\mathrm{d}^2I_{jk}}{\mathrm{d} t^2},
\end{equation}
where $I_{jk}=\sum_am_ax^j_ax^k_a$ is the mass quadrupole moment. As in GR, the TT part of $\tilde h_{jk}$ is also related to the reduced quadrupole moment $J_{jk}=I_{jk}-\delta_{jk}\delta^{il}I_{il}/3$,
\begin{equation}\label{hjkloj}
  \tilde h_{jk}^\mathrm{TT}=\frac{1}{8\pi G_{4(0,0)}r}\frac{\mathrm{d}^2J_{jk}^\mathrm{TT}}{\mathrm{d} t^2}.
\end{equation}

The leading order term for the scalar field $\varphi$ is the mass monopole which does not contribute to the effective stress-energy tensor, so it is necessary to take higher order terms into account.
To do so, the scalar equation (\ref{eq-phieomupto2}) is rewritten with the linearized equations  substituted in, which is given by
\begin{equation}\label{hhsclequ}
\begin{split}
  &(\Box-m_s^2)\varphi=\\
  &\frac{T_*^{(1)}}{2G_{4(0,0)}\zeta}+\frac{G_{4(1,0)}T^{(2)}}{2G_{4(0,0)}\zeta}-\frac{1}{\zeta}\left(\frac{\partial T}{\partial\phi}\right)^{(2)}\\
  &+\left[\frac{(T_*^{(1)})^2}{4G_{4(0,0)}^2\zeta^3}-\frac{(\partial_\mu\partial_\nu\varphi)(\partial^\mu\partial^\nu\varphi)}{\zeta}+\frac{m_s^2\varphi T_*^{(1)}}{G_{4(0,0)}\zeta^2}\right.\\
  &\left.+\frac{m_s^4\varphi^2}{\zeta}\right]\left(G_{3(0,1)}-3\frac{G_{4(0,1)}G_{4(1,0)}}{G_{4(0,0)}}+3\frac{G_{4(1,0)}G_{5(1,0)}}{G_{4(0,0)}}\right.\\
  &\left.-3G_{4(1,1)}\right)+\left[\frac{\varphi T_*^{(1)}}{G_{4(0,0)}\zeta}+2m_s^2\varphi^2+(\partial_\mu\varphi)(\partial^\mu\varphi)\right]\\
  &\left(-\frac{G_{4(1,0)}}{2G_{4(0,0)}}+\frac{3G_{4(1,0)}^3}{2G_{4(0,0)}^2\zeta}-\frac{G_{2(1,1)}}{2\zeta}+\frac{G_{3(2,0)}}{\zeta}\right.\\
  &\left.-3\frac{G_{4(1,0)}G_{4(2,0)}}{G_{4(0,0)}\zeta}\right)+\frac{G_{4(1,0)}}{G_{4(0,0)}}(\partial_\mu\varphi)\partial^\mu\varphi-\frac{\tilde hT_*^{(1)}}{4G_{4(0,0)}\zeta}\\
  &+\frac{T_{\mu\nu}^{(1)}\partial^\mu\partial^\nu\varphi}{G_{4(0,0)}\zeta}(G_{4(0,1)}-G_{5(1,0)})+\frac{\varphi T^{(1)}}{2G_{4(0,0)}\zeta}\left(G_{4(2,0)}\right.\\
  &\left.-\frac{G_{4(1,0)}^2}{G_{4(0,0)}}\right)-\varphi^2\left(\frac{G_{2(3,0)}}{2\zeta}+m_s^2\frac{G_{4(1,0)}}{G_{4(0,0)}}\right)\\
  &+\tilde h_{\mu\nu}\partial^\mu\partial^\nu\varphi-\frac{m_s^2\varphi\tilde h}{2}.
\end{split}
\end{equation}

In the following discussion, it is assumed that the scalar field is {\it massless} for simplicity. The details to obtain the following results can be found in \ref{app-pne}.
The leading order contribution to $\varphi$ comes from the first term on the right hand side of Eq.~(\ref{hhsclequ}), which is the mass monopole moment,
\begin{equation}\label{vpmono}
  \varphi^{[1]}=\frac{1}{8\pi G_{4(0,0)}\zeta r}\sum_am_aS_a.
\end{equation}
From now on, the superscript $[n]$ indicates the order of a quantity in terms of the speed $v$, i.e., $\varphi^{[n]}$ is of the order $O(v^{2n})$.
$\varphi^{[1]}$ is independent of time, so it does not contribute to the effective stress-energy tensor. The next leading order term is the mass dipole moment,
\begin{equation}\label{vpdip}
  \varphi^{[1.5]}=\frac{1}{8\pi G_{4(0,0)}\zeta r}\sum_am_aS_a(\hat n\cdot\vec v_a),
\end{equation}
in which $\hat n=\vec x/r$.
This gives the leading contribution to the effective stress-energy tensor.
At the next next leading order, there are more contributions from the remaining terms on the right hand side of Eq.~(\ref{hhsclequ}). First, there is the mass quadruple moment contribution,
\begin{equation}\label{vpmquad}
     \varphi_1^{[2]}=\frac{1}{8\pi G_{4(0,0)}\zeta r}\sum_am_aS_a[(\hat n\cdot\vec a_a)(\hat n\cdot\vec x_a)+(\hat n\cdot\vec v_a)^2].
\end{equation}
And the remaining contribution to the scalar wave is
\begin{equation}\label{varphi2}
\begin{split}
  &\varphi_2^{[2]}=\\
  &-\frac{1}{16\pi G_{4(0,0)}\zeta r}\sum_am_aS_av_a^2\\
  &+\frac{1}{64\pi^2G_{4(0,0)}^2\zeta r}\sideset{}{'}\sum_{a,b}\frac{m_am_b}{r_{ab}}\left(-\frac{S_a}{2}\right.\\
  &\left.+\frac{3G_{4(1,0)}}{2G_{4(0,0)}\zeta}S_aS_b+\frac{S_a'S_b}{\phi_0\zeta}\right)\\
  &+\frac{1}{64\pi^2G_{4(0,0)}^2\zeta^2r}\left(G_{4(2,0)}-\frac{G_{4(1,0)}^2}{G_{4(0,0)}}\right)\sideset{}{'}\sum_{a,b}\frac{m_am_bS_b}{r_{ab}},\\
  &-\frac{G_{2(3,0)}}{256\pi^2G_{4(0,0)}^2\zeta^3r}\sideset{}{'}\sum_{a,b}m_am_bS_aS_br_{ab}\\
  &+\frac{1}{64\pi^2G_{4(0,0)}^2\zeta^2 r}\Upsilon\sideset{}{'}\sum_{a,b}\frac{m_am_bS_aS_b}{r_{ab}},
  \end{split}
\end{equation}
where $\sideset{}{'}\sum_{a,b}$ means summation over $a$ and $b$ with $a\ne b$, and
\[
\begin{split}
\Upsilon=&-\frac{3G_{4(1,0)}}{2G_{4(0,0)}}+\frac{3G_{4(1,0)}^3}{2G_{4(0,0)}^2\zeta}-\frac{G_{2(1,1)}}{2\zeta}+\frac{G_{3(2,0)}}{\zeta}\\
&-3\frac{G_{4(1,0)}G_{4(2,0)}}{G_{4(0,0)}\zeta}.
\end{split}
\]
Note that the penultimate line of Eq.~\eqref{varphi2} is a sum of terms proportional to $r_{ab}$, which grows as $r_{ab}$ increases and potentially dominates over other terms.
Since matters are confined within the source zone,
this line never blows up.

The scalar field up to the fourth order in velocity is given by
\begin{equation}\label{scf}
  \varphi=\varphi^{[1]}+\varphi^{[1.5]}+\varphi^{[2]}_1+\varphi^{[2]}_2.
\end{equation}
It is easy to check that this result agrees with Eq.~(86) in Ref.~\cite{Alsing:2011er} with $m_s=0$.

\section{Gravitational Radiation for a Compact Binary System}\label{secgwb}

This section is devoted to calculating the gravitational radiation for a compact binary system in the case with \emph{massless} scalar field .
According to Eq.~(\ref{effsettt}), the energy carried away by the gravitational wave is at a rate of
\begin{equation}\label{gwrad}
\begin{split}
  \dot E=&\oint T_{0j}^\mathrm{GW}\mathrm{d} S^j\\
  \approx&-\frac{G_{4(0,0)}}{2}r^2\int \left\langle\partial_0\tilde h_{jk}^\mathrm{TT}\partial_0\tilde h^{jk}_\mathrm{TT}\right\rangle\mathrm{d}\Omega\\
  &-\zeta r^2\int \left\langle\partial_0\varphi\partial_0\varphi\right\rangle\mathrm{d}\Omega,
\end{split}
\end{equation}
where the integration is carried out on a 2-sphere in the far zone
and in the final step, higher order terms have been dropped.
The first term gives the contribution of the spin-2 gravitational wave, while the second one gives the contribution of the scalar field.

Next, one has to calculate the motion of the binary system explicitly. By Eq.~(\ref{relaccb}), the relative acceleration is given by
\begin{equation}\label{relaccb1}
  a^j=-\frac{\varsigma m}{16\pi G_{4(0,0)}}\frac{\hat r^j_{12}}{r_{12}^2},
\end{equation}
where
 \begin{equation}\label{defdel}
   \varsigma=1+\frac{S_1S_2}{G_{4(0,0)}\zeta}.
 \end{equation}
As in GR, one can orient the coordinate system such that the orbit lies in the $xOy$ plane.
In the polar coordinate system $(r,\theta, z)$, the relative distance is thus given by
\begin{equation}\label{kerpleror}
  r_{12}(t)=\frac{p}{1+e\cos\theta(t)},
\end{equation}
where
\begin{equation}\label{defp}
  p=\frac{16\pi G_{4(0,0)}l^2}{\varsigma m},
\end{equation}
with $l$ the angular momentum per unit mass and $e$ the eccentricity.
The orbital period is
\begin{equation}\label{orpe}
  T=2\pi\sqrt{\frac{16\pi G_{4(0,0)}a^3}{\varsigma m}}.
\end{equation}
All these above results can be obtained by suitably modifying those in GR as found in Ref.~\cite{Poisson2014}.
Using Eq.~(\ref{mhamb}) with $m_s$ set to 0, the total mechanical energy of the binary system is
\begin{equation}\label{totmee}
  E=-\frac{\varsigma\mu m}{32\pi G_{4(0,0)}a},
\end{equation}
where $a=p/(1-e^2)$ is the semi-major axis.

Following Ref.~\cite{Alsing:2011er}, the rate of energy loss due to the spin-2 gravitational wave is
\begin{equation}\label{gwradtene}
  \dot E_2=-(1-e^2)^{-7/2}\left(1+\frac{73}{24}e^2+\frac{37}{96}e^4\right)\frac{32}{5}\frac{\varsigma^3\mu^2m^3}{(16\pi G_{4(0,0)})^4a^5},
\end{equation}
which reproduces the radiation damping of GR in the appropriate limit \cite{Poisson2014}.

Ignoring the leading order contribution to $\varphi$, the higher order correction is given by
\begin{equation}\label{hoss}
\begin{split}
 \varphi =&\frac{f_1}{r}(\hat n\cdot \vec v)+\frac{f_2}{r}(\hat n\cdot\vec v)^2+\frac{f_3}{r}\frac{(\hat n\cdot\vec r_{12})^2}{r_{12}^3}\\
 &+\frac{f_4}{r}v^2+\frac{f_5}{rr_{12}}+f_6\frac{r_{12}}{r},
 \end{split}
\end{equation}
where
\begin{gather}
f_1=-\frac{\mu}{4\pi \phi_0\zeta}(s_1-s_2),\quad f_2=\frac{\mu \Gamma}{8\pi G_{4(0,0)}\zeta},\\
f_3=-\frac{\varsigma\mu m\Gamma}{128\pi^2 G_{4(0,0)}^2\zeta},\quad f_4=-\frac{\mu\Gamma}{16\pi G_{4(0,0)}\zeta},\\
f_5=-\frac{\mu m\Gamma'}{64\pi^2G_{4(0,0)}^2\zeta}+\frac{\mu m\Gamma'}{32\pi^2G_{4(0,0)}^2\zeta^2}\left(G_{4(2,0)}-\frac{G_{4(1,0)}^2}{G_{4(0,0)}}\right)\nonumber\\
  +\frac{\mu m}{64\pi^2G_{4(0,0)}^2\zeta^2}\Bigg[\left(\frac{3G_{4(1,0)}^3}{2G_{4(0,0)}^2}-\frac{G_{2(1,1)}}{2}+G_{3(2,0)}\right.\nonumber\\
  \left.-\frac{3G_{4(1,0)}G_{4(2,0)}}{G_{4(0,0)}}\right) \frac{2S_1S_2}{\zeta}+\frac{S'_1S_2+S'_2S_1}{\phi_0}\Bigg],\\
f_6=-\frac{\mu mG_{2(3,0)}S_1S_2}{128\pi^2G_{4(0,0)}^2\zeta^3},
\end{gather}
and
\begin{gather}
S'_a=G_{4(1,0)}s_a-\frac{2G_{4(0,0)}}{\phi_0}(s_a^2-s_a-s'_a),\\
  \Gamma=G_{4(1,0)}-\frac{2G_{4(0,0)}}{\phi_0}\frac{m_2s_1+m_1s_2}{m},\\
  \Gamma'=G_{4(1,0)}-\frac{G_{4(0,0)}}{\phi_0}(s_1+s_2).
\end{gather}
The first term at the right hand side of Eq.~(\ref{hoss}) is a dipolar contribution and oscillates at the orbital frequency. This term is of order $v^{-1}\gg1$ relative to the remaining terms.
However, it also depends on the difference in the sensitivities $(s_1-s_2)$ of the objects in the binary system, which might be small or even vanish.
For example, in the Shift-Symmetric Horndeski theory (SSHT) with $G_i$  functions of $X$ only, the stellar sensitivity $s_a$ vanishes \cite{Barausse:2015wia},
and in Brans-Dicke theory, the sensitivity of a black hole is 1/2 \cite{Hawking:1972qk,Zaglauer1992,Alsing:2011er}.
So if the binary system consists of, e.g., two neutron stars in SSHT or if the two stars are black holes in Brans-Dicke theory, the dipolar radiation vanishes.

In the generic case, $(s_1-s_2)$ might not be zero, and the dipolar contribution should be taken into account.
So the contribution of the scalar field to the energy flux is
\begin{equation}\label{gwradscle}
  \begin{split}
\dot E_0=&-\zeta r^2\int \left\langle\partial_0\varphi\partial_0\varphi\right\rangle\mathrm{d}\Omega\\
        =&-(1-e^2)^{-7/2}\times\\
        &\Bigg\{\frac{\zeta\varsigma^3 m^3}{120(16\pi)^2G_{4(0,0)}^3a^5}\Big[15 (e^2+4) e^2 f_4^2\\
        &+10 (e^2+4) e^2 f_2 f_4+(6 e^4+36 e^2+8) f_2^2\Big]\\
        &+\frac{\zeta\varsigma^2 m^2}{1920\pi G_{4(0,0)}^2a^5}\Big[-5a(1-e^2)(2+e^2)f_1^2 \\
        &+(3 e^4+36 e^2+16) f_2f_3-5 (e^2+4) e^2 f_2f_5\\
        & + 20 a^2e^2(1-e^2)^2 f_2f_6-5e^2(e^2+4) f_3 f_4\\
        &-15e^2(e^2+4)f_4f_5+60a^2e^2(1-e^2)^2f_4f_6\Big]\\
        &+\frac{\zeta \varsigma m}{480G_{4(0,0)}a^5}\Big[(15 e^4+108 e^2+32) f_3^2\\
        &+15 e^2 (e^2+4) f_5^2+10 e^2 (e^2+4)f_3  f_5\\
        &-120 a^4(1-1/\sqrt{1-e^2})(1-e^2)^4f_6^2\\
        & -120 a^2e^2(1-e^2)^2  f_5 f_6\\
        & -40a^2 e^2(1-e^2)^2f_3f_6 \Big]\Bigg\}.
\end{split}
\end{equation}
A straightforward but tedious calculation shows that Eq.~(\ref{gwradscle}) reduces to Eq.~(3.24) in Ref.~\cite{Brunetti:1998cc} for Brans-Dicke theory with sensitivities set to zero and the Hadamard regularization imposed \cite{Hadamard1923bk,Sellier1994,Blanchet:2000nu}.
The period change $\dot T$ can be measured experimentally, and the fractional period change $\dot T/T$ is given by
\begin{equation}\label{derpere}
  \begin{split}
     \frac{\dot T}{T}=&-\frac{3}{2}\frac{\dot E_0+\dot E_2}{E}\\
     =&-(1-e^2)^{-7/2}\times\\
     &\Bigg\{\frac{96}{5}\left(1+\frac{73}{24}e^2+\frac{37}{96}e^4\right)\frac{\varsigma^2\mu m^2}{(16\pi G_{4(0,0)})^3a^4}\\
     &+\frac{\zeta\varsigma^2m^2}{640\pi\mu G_{4(0,0)}^2a^4}\Big[15 (e^2+4) e^2 f_4^2\\
     &+10 (e^2+4) e^2 f_2 f_4+(6 e^4+36 e^2+8) f_2^2\Big]\\
&+\frac{\zeta\varsigma m}{40\mu G_{4(0,0)}a^4}\Big[-5a(1-e^2)(2+e^2)f_1^2 \\
&+(3 e^4+36 e^2+16) f_2f_3-5 (e^2+4) e^2 f_2f_5 \\
&+ 20 a^2e^2(1-e^2)^2 f_2f_6-5e^2(e^2+4) f_3 f_4\\
&-15e^2(e^2+4)f_4f_5+60a^2e^2(1-e^2)^2f_4f_6\Big]\\
&+\frac{\pi\zeta}{10\mu a^4}\Big[(15 e^4+108 e^2+32) f_3^2\\
   &+15 e^2 (e^2+4) f_5^2+10 e^2 (e^2+4)f_3  f_5\\
   &-120 a^4(1-1/\sqrt{1-e^2})(1-e^2)^4 f_6^2\\
   & -120 a^2e^2 (1-e^2)^2 f_5 f_6 \\
   &-40a^2 e^2(1-e^2)^2f_3f_6 \Big]\Bigg\}.
  \end{split}
\end{equation}
The first term is caused by the spin-2 gravitational wave, while the remaining ones by the scalar field.

Given the sensitivities ($s_a, s'_a$) of all kinds of celestial objects, Eq.~(\ref{derpere}) can be compared with the observed period change to set bounds on some of parameters characterizing a particular scalar-tensor theory (e.g., $\phi_0,\,G_{4(0,0)},\,G_{4(1,0)},\,\zeta$ etc.) as done in Ref.~\cite{Alsing:2011er}.

\section{Observational Constraints}\label{sec-exp}

In this section, constraints on Horndeski theory are obtained using observations from lunar laser ranging experiments, Cassini time-delay measurement and  binary pulsars.
Since Horndeski theory contains many parameters, the following discussions start with generic constraints on the full Horndeski theory, and then specify to some concrete subclasses of Horndeski theory.

\subsection{Constraints from lunar laser ranging experiments}\label{sec-llr}

The lunar laser ranging experiment gave the most precise measurement of the Nordtvedt effect, and the Nordtvedt parameter was determined to be \cite{Hofmann2010llr}
\begin{equation}\label{nordpar}
  \eta_\text{N}^\text{obs.}=(0.6\pm5.2)\times10^{-4}=\delta_1\pm\epsilon_1.
\end{equation}
To get the constraints, one requires that $|\eta_\text{N}-\delta_1|<2\epsilon_1$ at 95\% confidential level. Using Eq.~(\ref{effNordpar}), one obtains
\begin{equation}\label{consnord}
  -0.98\times10^{-3}<\frac{G_{4(1,0)}(1+m_sr)}{8\pi G_\text{N} G_{4(0,0)}\phi_0\zeta}e^{-m_sr}<1.1\times10^{-3},
\end{equation}
where $r=1$ AU and the sensitivity of the Sun is ignored as its sensitivity is expected to be smaller than $10^{-4}$, which is the white dwarf's sensitivity \cite{Zaglauer1992,Alsing:2011er}.

\subsection{Constraints from Cassini time-delay data}\label{sec-cas}

In 2002, the Cassini spacecraft measured the Shapiro time delay effect in the solar system by radio tracking \cite{Bertotti:2003rm}.
The PPN parameter $\gamma$ was given by
\begin{equation}\label{mesga}
  \gamma_\text{meas.}=1+(2.1\pm2.3)\times10^{-5}=1+\delta_2\pm\epsilon_2.
\end{equation}
At 95\% confidential level, one requires that $|\gamma(r)-\gamma_\text{meas.}|<2\epsilon_2$, which leads to
\begin{equation}\label{consstd}
  -3.35\times10^{-5}\lesssim\frac{G_{4(1,0)}^2}{G_{4(0,0)}\zeta}e^{-m_sr}\lesssim1.25\times10^{-5},
\end{equation}
in which the Sun's sensitivity is also ignored, and $r=1$ AU.
In the massless case, this constraint can be translated into $\omega_\text{H}\gtrsim4\times10^4$ with $\omega_\text{H}=G_{4(0,0)}(G_{2(0,1)}-2G_{3(1,0)})/2G_{4(1,0)}^2$ \cite{Hohmann:2015kra}, which reduces to $\omega_\text{BD}$ when (the massless) Brans-Dicke theory is considered.

\subsection{Constraints from period change for circular motion}

Now, one obtains the constraints on Horndeski theory using the data of pulsars. For this end, one considers the circular motion of a binary system, not only for simplicity but also because
the first sensitivities $s_a$ are known at least in some subclasses of Horndeski theory, such as Brans-Dicke theory \cite{Hawking:1972qk,Zaglauer1992,Alsing:2011er} and SSHT \cite{Barausse:2015wia}, while the second sensitivities $s'_a$ are unknown.
In the case of the circular motion ($e=0$), one assumes that $\omega$ is the orbital angular frequency so that $r_{12}=a$ and $\theta=\omega t$.
The orbital angular frequency can be obtained using Eq.~(\ref{defp}), which is
\begin{equation}\label{orfreq}
  \omega=\frac{2\pi}{T}=\sqrt{\frac{\varsigma m}{16\pi G_{4(0,0)}r_{12}^3}}.
\end{equation}
The total mechanical energy of the binary system is
\begin{equation}\label{totme}
  E=-\frac{\varsigma \mu m}{32\pi G_{4(0,0)}r_{12}}.
\end{equation}
The rates of radiation damping are greatly simplified,
\begin{equation}\label{gwradten}
  \dot E_2=-\frac{32}{5}\frac{\varsigma^3\mu^2m^3}{(16\pi G_{4(0,0)})^4r_{12}^5},
\end{equation}
and
\begin{equation}\label{gwradscl}
  \dot E_0=-\frac{1}{12\pi}\frac{\varsigma^2\mu^2m^2(s_1-s_2)^2}{(16\pi G_{4(0,0)})^2\phi_0^2\zeta r_{12}^4}-\frac{16}{15}\frac{\varsigma^3\mu^2m^3\Gamma^2}{(16\pi)^4G_{4(0,0)}^5\zeta r_{12}^5},
\end{equation}
where the first term comes from the mass dipole moment. The fractional period change is
\begin{equation}\label{rcop}
\begin{split}
  \frac{\dot T}{T}=&-\frac{\varsigma\mu m(s_1-s_2)^2}{64\pi^2 G_{4(0,0)}\phi_0^2\zeta r_{12}^3}-\frac{16}{5}\frac{\varsigma^2\mu m^2\Gamma^2}{(16\pi)^3 G_{4(0,0)}^4\zeta r_{12}^4}\\
  &-\frac{96}{5}\frac{\varsigma^2\mu m^2}{(16\pi G_{4(0,0)})^3r_{12}^4}.
  \end{split}
\end{equation}
The first two terms are caused by the scalar field, while the last one by the spin-2 gravitational wave.

Provided that the sensitivities ($s_1,s_2$) of  celestial objects are given, Eq.~(\ref{rcop}) can be compared with the observed period change to set bounds on some  parameters in Horndeski theory,
using the observational data of the binary system PSR J1738+0333 \cite{Freire:2012mg}.
This is a 5.85-ms pulsar with a white dwarf companion, orbiting around each other every 8.51 hours.
Some of the orbit parameters are tabulated in Table \ref{tab-psrj1739+0333}.
\begin{table}[h]
  \centering
   \caption{Orbital parameters of the binary system PSR J1738+0333 \cite{Freire:2012mg}.}\label{tab-psrj1739+0333}
  \begin{tabular}{lc}
    \hline
    \hline
    Eccentricity $e$ & $(3.4\pm1.1)\times10^{-7}$ \\
    Orbital period $T$ (days) & 0.354 790 739 8724(13) \\
    Period change $\dot T_\text{obs}$ & $(-25.9\pm3.2)\times10^{-15}$ \\
    Pulsar mass $m_1(M_\odot)$ & $1.46_{-0.05}^{+0.06}$ \\
    Companion mass $m_2(M_\odot)$ & $0.181_{-0.007}^{+0.008}$ \\
    \hline
  \end{tabular}
\end{table}
The eccentricity of PSR J1738+0333 is $(3.4\pm1.1)\times10^{-7}$, so the orbit is nearly a circle, and one can use Eq.~(\ref{rcop}) to obtain the bounds on Horndeski theory.
At 95\% confidential level, one requires that $|\dot T_\text{pred.}-\dot T_\text{obs.}|<2\sigma$ where $\dot T_\text{pred.}$ is determined by Eq.~(\ref{rcop}) with Eq.~\eqref{orfreq} substituted in, $\dot T_\text{obs.}$ is the observed period change and $\sigma$ is the uncertainty for $\dot T_{\text{obs.}}$.
The expression for $\dot T_\text{pred.}-\dot T_\text{obs.}$ is too complicated and will not be presented here.

\subsection{Constraints on Special Examples}

\begin{figure}

  \includegraphics[width=0.5\textwidth]{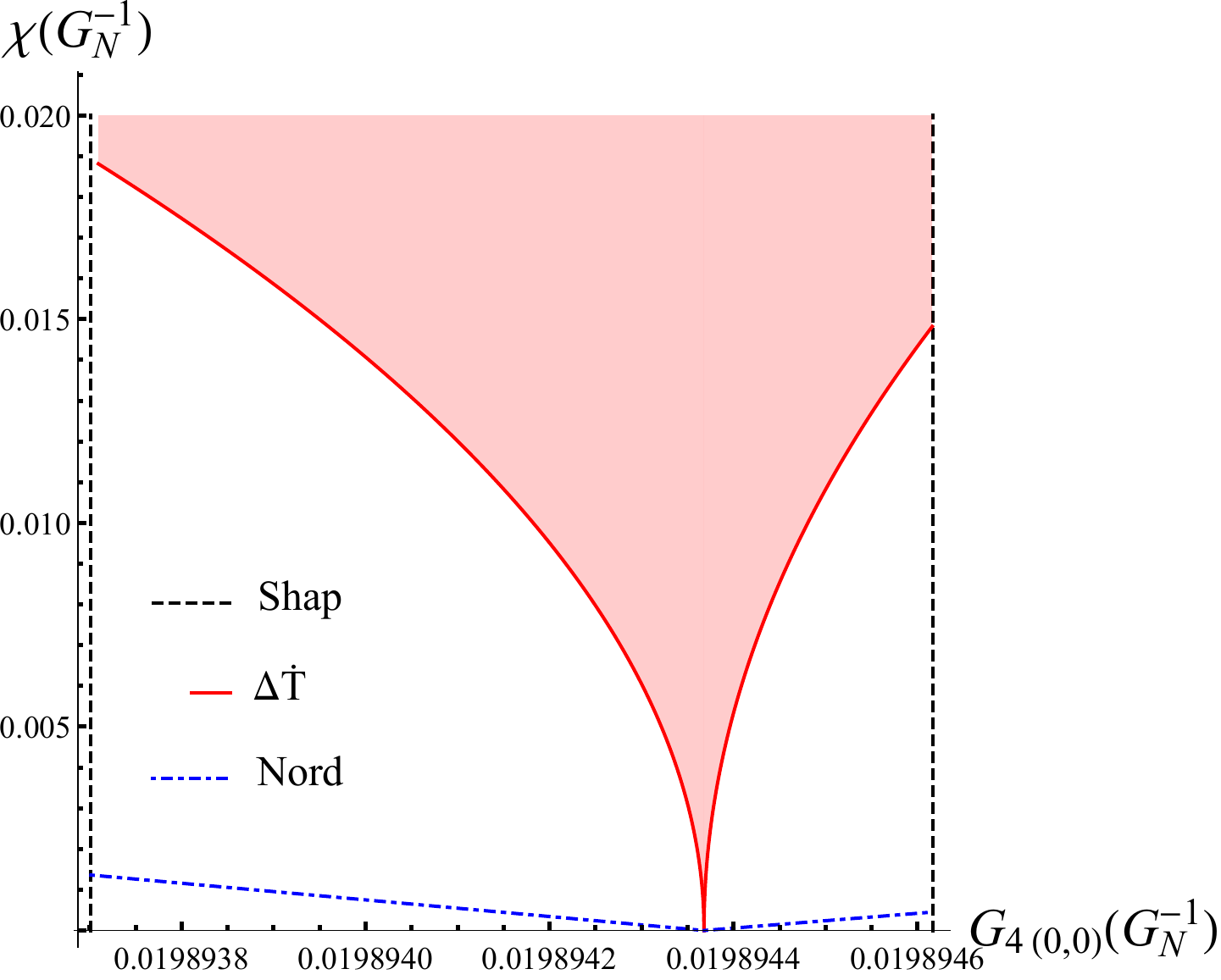}
  \caption{The allowed parameter spaces $(G_{4(0,0)},\chi)$ set by the Nordtvedt effect (the region above the dot dashed blue curve), the Shapiro time delay (the region enclosed by the two vertical, dashed black lines) and the observation of the binary pulsar PSR J1738+0333 (the region above the solid red curve, labeled by $\Delta \dot T$), respectively.
  The shaded area is the commonly allowed parameter space.
  The horizontal and the vertical axes are both measured in units of $G^{-1}_\text{N}$.
  }\label{fig-cons-ml}
\end{figure}

{\it Example 1}: Consider a special subclass of Horndeski theory where the scalar field is massless, i.e., $G_{2(2,0)}=0$.
By Eq.~\eqref{eq-gn-ml}, one can solve for $\zeta$ in terms of $G_{4(0,0)}$ and $G_{4(1,0)}$,
\begin{equation}\label{eq-zeta-ml}
  \zeta=\frac{G_{4(1,0)}^2}{G_{4(0,0)}(16\pi G_{4(0,0)}G_\text{N}-1)}.
\end{equation}
Note that since the Newton's constant $G_\text{N}$ is measured in the vicinity of the Earth, the Earth's sensitivity $s_\otimes$ is ignored in Eq.~\eqref{eq-gn-ml}, and so $\zeta$ does not depend on $s_\otimes$.
Plug $\zeta$ into Eq.~\eqref{consstd}, and the Shapiro time delay effect constrains $G_{4(0,0)}$,
\begin{equation}\label{eq-g400-c-ml}
  \frac{1-3.35\times10^{-5}}{16\pi G_\text{N}}\lesssim G_{4(0,0)}\lesssim\frac{1+1.25\times10^{-5}}{16\pi G_\text{N}}.
\end{equation}
Plug $\zeta$ into Eq.~\eqref{consnord}, and one gets
\begin{equation}\label{eq-g410-c-ml}
  -0.98\times10^{-3}\lesssim\frac{16\pi G_{4(0,0)}G_\text{N}-1}{8\pi\phi_0G_{4(1,0)}G_\text{N}}\lesssim1.1\times10^{-3},
\end{equation}
which shows a nice property that the product $\chi=\phi_0G_{4(1,0)}$ appears in the above expression.
In fact, after one substitutes $\zeta$ into Eq.~\eqref{rcop}, $\dot T$ can also be expressed as a function of $G_{4(0,0)}$ and $\chi$, which is too complicated to be presented.
Note that the sensitivities for the pulsar and the white dwarf are taken to be approximately 0.2 and $10^{-4}$, respectively.
So the constraints from the Nordtvedt effect and the period change of the binary pulsar can be represented by the constraints on $G_{4(0,0)}$ and $\chi$.
The result is given in Fig.~\ref{fig-cons-ml}.
The shaded area is the commonly allowed parameter space $(G_{4(0,0)},\chi)$.
Finally, since $\zeta$ is given in Eq.~\eqref{msq}, one knows that
\begin{equation}\label{eq-g2g3-ml}
\begin{split}
 & \frac{G_{2(0,1)}-2G_{3(1,0)}}{G_{4(1,0)}^2}\gtrsim4.02\times10^6G_\text{N}\,\text{or } \\ &\frac{G_{2(0,1)}-2G_{3(1,0)}}{G_{4(1,0)}^2}\lesssim-1.50\times10^6G_\text{N}.
\end{split}
\end{equation}
Note that the above constraints cannot be applied to the special case where $G_4\propto\phi$, as in this case, $G_{4(1,0)}\propto G_{4(0,0)}/\phi_0$, i.e., $G_{4(1,0)}$ and $G_{4(0,0)}$ are not independent of each other.

{\it Example 2}:
Now, consider a second subclass of Horndeski theory whose $G_4=G_4(\phi)$ and $G_5=0$.
The scalar field is still assumed to be massless.
This subclass satisfies the constraints set by the gravitational wave speed limit \cite{Creminelli:2017sry,Ezquiaga:2017ekz,Baker:2017hug}.
One can introduce a new scalar field $\phi'$ such that $G_4(\phi)=\phi'/16\pi$, and the form of action \eqref{acth} remains the same after replacing $\phi$ by $\phi'$ in it.
So let us simply call the new scalar field $\phi$, and thus $G_4(\phi)=\phi/16\pi$ and $G_{4(1,0)}=1/16\pi$.
Using all the constraints discussed in the previous subsections, one obtains
\begin{equation}\label{eq-phi-ml-s}
  \frac{1-3.35\times10^{-5}}{G_\text{N}}\lesssim \phi_0\lesssim\frac{1+1.25\times10^{-5}}{G_\text{N}},
\end{equation}
and this leads to
\begin{equation}\label{eq-g2g3-ml-s}
  G_{2(0,1)}-2G_{3(1,0)}\gtrsim 1600G_\text{N}\,\text{or } G_{2(0,1)}-2G_{3(1,0)}\lesssim-600G_\text{N}.
\end{equation}


{\it Example 3}:
One may also consider the constraints set on a massive Horndeski theory.
In this case, one can only use the constraints from the Nordtvedt effect and the Shapiro time delay.
The mass $m_s$ of the scalar field is expected to be very small.
As suggested in Ref.~\cite{Alsing:2011er}, if $10^{-21}\text{ eV}<m_s<10^{-15}$ eV, the constraints can also be set on $G_{4(0,0)}$ and $\chi$, provided that they are independent of each other. The allowed parameter space $(G_{4(0,0)},\chi)$ is approximately given by the area enclosed by the two vertical dashed curves, and the dot dashed one in Fig.~\ref{fig-cons-ml}.
The constraint on the combination $G_{2(0,1)}-2G_{3(1,0)}$ is also approximately given by Eq.~\eqref{eq-g2g3-ml}.
If $G_4\propto\phi$, the constraints are approximately given by Eqs.~\eqref{eq-phi-ml-s} and \eqref{eq-g2g3-ml-s}.

\section{Conclusion}\label{seccon}

In this work, the observational constraints on Horndeski theory are obtained based on the observations from the Nordtvedt effect, Shapiro time delay and binary pulsars.
For this purpose, the near zone metric and scalar perturbations are first calculated in order to obtain the equations of motion for the stars.
These solutions are thus used to study the Nordtvedt effect and the Shapiro time delay.
Then, the effective stress-energy tensor of Horndeski theory is derived using the method of Isaacson.
It is then used to calculate the rate of energy radiated away by the gravitational wave and the period change of a binary system.
For this end, in the far zone, the auxiliary metric perturbation is calculated using the familiar quadratic formula, and the scalar field is calculated with the monopole moment contribution dominating, although it does not contribute to the effective stress-energy tensor.
The leading contribution of the scalar field to the energy damping is the dipolar radiation, which is related to the difference in the sensitivities of the stars in the binary system, so the dipolar radiation vanishes
if the two stars have the same sensitivity.
The energy damping is finally calculated with the far zone field perturbations, and the period change is derived.
Finally, the observational constraints are discussed based on the data from lunar laser ranging experiments, the observations made by the Cassini spacecraft,  and the observation on the PSR J1738+0333.
Explicit constraints have been obtained for both the massless and massive Horndeski theory, and in particular, for the one satisfying the recent gravitational wave speed limits \cite{TheLIGOScientific:2017qsa}.

\begin{acknowledgements}
We would like to thank Zhoujian Cao for helpful discussions.
This research was supported in part by the Major Program of the National Natural Science Foundation of China under Grant No. 11475065 and the National Natural Science Foundation of China under Grant No. 11690021.
\end{acknowledgements}

\appendix

\section{Equations of Motion up to the Second Order in Perturbations}\label{app-eoms2nd}

In this work, the equations of motion are obtained and simplified using {\it xAct} package \cite{Martin-Garcia:2007bqa,MartinGarcia:2008qz,xperm2008,Brizuela:2008ra,xact}. The equations agree with those listed in Refs.~\cite{Kobayashi:2011nu,Gao:2011mz}.
The equations are then perturbed around a generic background spacetime up to second order in perturbations in $g_{\mu\nu}$ and $\phi$.
Finally, the background spacetime is set to be Minkowskian,
and the resultant equations of motion up to the second order in perturbations are
\begin{eqnarray}
   &&\frac{1}{2}(T_{\mu\nu}^{(1)}+T_{\mu\nu}^{(2)})\nonumber\\
   &=&-\frac{G_{2(0,0)}}{2}\eta_{\mu\nu}+G_{4(0,0)}(G_{\mu\nu}^{(1)}+G_{\mu\nu}^{(2)})-G_{4(1,0)}(\partial_\mu\partial_\nu\varphi\nonumber\\
   &&-\eta_{\mu\nu}\Box\varphi)+G_{4(1,0)}\left( R^{(1)}_{\mu \nu} \varphi  -  \frac{1}{2} \eta_{\mu \nu} R^{(1)} \varphi \right) \nonumber\\
   &&+  (G_{4(0,1)}-G_{5(1,0)})\Big[(\partial_{\mu}\partial^{\rho}\varphi)\partial_{\nu}\partial_{\rho}\varphi -  (\partial_{\mu}\partial_{\nu}\varphi) \Box\varphi\nonumber\\
   && + \frac{1}{2} \eta_{\mu \nu} (\Box\varphi)^2 - \frac{1}{2} (\partial_{\rho}\partial_{\sigma}\varphi) \partial^{\rho}\partial^{\sigma}\varphi \Big]\nonumber\\
   &&+\left(G_{3(1,0)}-\frac{G_{2(0,1)}}{2}\right)\left[(\partial_{\mu}\varphi) \partial_{\nu}\varphi-\frac{1}{2} \eta_{\mu \nu} \Box\varphi \right]\nonumber\\
   &&+G_{4(1,0)}\left\{ \frac{1}{2} (\partial^{\rho}\varphi) \partial_{\mu}h_{\nu \rho}  + \frac{1}{2} (\partial^{\rho}\varphi) \partial_{\nu}h_{\mu \rho}    \right. \nonumber\\
   &&\left.-  \frac{1}{2} (\partial^{\rho}\varphi) \partial_{\rho}h_{\mu \nu} + h_{\mu \nu} \Box\varphi +\eta_{\mu \nu}\left[\frac{1}{2}  (\partial_{\rho}h) \partial^{\rho}\varphi  \right.\right.\nonumber\\
    &&\left.\left.-   (\partial^{\rho}\varphi) \partial_{\sigma}h_{\rho}^{\sigma}  -   h_{\rho\sigma} \partial^{\rho}\partial^{\sigma}\varphi\right]\right\}\nonumber \\
   &&+ G_{4(2,0)} \{\eta_{\mu \nu} [(\partial_{\rho}\varphi) \partial^{\rho}\varphi  + \varphi \Box\varphi] \nonumber\\
   && -  (\partial_{\mu}\varphi) \partial_{\nu}\varphi  -  \varphi \partial_{\mu}\partial_{\nu}\varphi \}- \frac{1}{4}G_{2(2,0)} \eta_{\mu \nu} \varphi^2,\label{eq-eineomupto2}\\
 && -\left(\frac{\partial T}{\partial \phi}\right)^{(1)}-\left(\frac{\partial T}{\partial \phi}\right)^{(2)}\nonumber\\
 &=&G_{2(1,0)}+(G_{2(0,1)}-2G_{3(1,0)})\Box\varphi\nonumber\\
  &&+(G_{2(2,0)}+G_{4(1,0)} )(R^{(1)}+R^{(2)})+G_{4(2,0)}\varphi R^{(1)} \nonumber\\
  &&+(G_{4(0,1)}-G_{5(1,0)})( R^{(1)} \Box\varphi - 2 R^{(1)}_{\mu\nu} \partial^{\mu}\partial^{\nu}\varphi )\nonumber \\
  && +G_{2(0,1)}\left[  \frac{1}{2} (\partial_{\nu}\varphi) \partial^{\nu}h  -  h^{\mu\nu} \partial_{\mu}\partial_{\nu}\varphi -  (\partial_{\mu}h^{\mu\nu}) \partial_{\nu}\varphi \right] \nonumber\\
  && + (G_{3(0,1)}-3G_{4(1,1)})[(\partial_{\mu}\partial_{\nu}\varphi) \partial^{\mu}\partial^{\nu}\varphi  -  (\Box\varphi)^2]\nonumber \\
  &&+G_{3(1,0)}[2 h^{\mu\nu} \partial_{\mu}\partial_{\nu}\varphi  + 2 (\partial_{\mu}h^{\mu\nu}) \partial_{\nu}\varphi -  (\partial_{\nu}\varphi) \partial^{\nu}h ]  \nonumber\\
  &&+ (G_{2(1,1)}-2 G_{3(2,0)})\left[ \frac{1}{2}  (\partial_{\mu}\varphi) \partial^{\mu}\varphi  + \varphi \Box\varphi\right]\nonumber\\
  &&   + \frac{1}{2}  G_{2(3,0)}\varphi^2,\label{eq-phieomupto2}
\end{eqnarray}
where $\Box=\eta^{\mu\nu}\partial_\mu\partial_\nu$, and the superscript $(1)$ implies the leading order piece of the quantity while the superscript $(2)$ represents the second order piece.

\section{Post-Newtonian Expansion of the Scalar Field}\label{app-pne}

In this appendix, the procedure to derive the post-Newtonian expansion of the scalar field is briefly presented.
The basic idea is the following.
Suppose a scalar field $\psi$ satisfies the massless Klein-Gordon equation with a source $S$,
\begin{equation}\label{kgsc}
  \Box\psi=-16\pi S,
\end{equation}
where $\Box=\partial_\mu\partial^\mu$. In the far zone, the scalar field is given by
\begin{equation}\label{fzsc}
  \psi(t,\vec x)=4\int_{\mathcal N}\frac{S(t-|\vec x-\vec x'|,\vec x')}{|\vec x-\vec x'|}\mathrm{d}^3x'.
\end{equation}
Here, the integration is over the near zone $\mathcal{N}$,
as $\psi$ will be calculated only up to the quadratic order in perturbations.
Since $r=|\vec x|>|\vec x'|$, one can expand the integrand in powers of $\vec x'$ in the following way,
\begin{equation}\label{fzexp}
  \psi(t,\vec x)=4\sum_{q=0}^\infty\frac{(-1)^q}{q!}\partial_Q\left(\frac{I^Q(u)}{r}\right),
\end{equation}
where $u=t-r$ is the retarded time, $Q$ is a multi-index, namely, $\partial_Q=\partial_{j_1}\partial_{j_2}\cdots\partial_{j_q}$ and $I^Q=I^{j_1j_2\cdots j_q}$, and the repeated indices imply summation. The symbol $I^Q(u)$ is
\begin{equation}\label{defiq}
  I^Q(u)=\int_\mathcal{M}S(u,\vec x')x'^Q\mathrm{d}^3x',
\end{equation}
in which the integration is over $\mathcal{M}$, the intersection of the near-zone worldtube with the constant retarded time hypersurface $u=C$.
Since $\partial_ju=-x_j/r=-\hat n_j$, Eq.~(\ref{fzexp}) is approximately given by
\begin{equation}\label{fzexpr}
  \psi(t,\vec x)=\frac{4}{r}\sum_{q=0}^{\infty}\frac{1}{q!}\frac{\partial^q}{\partial t^q}\int_\mathcal{M}S(u,\vec x')(\hat n\cdot\vec x')^q\mathrm{d}^3x'+O(1/r^2).
\end{equation}
For the purpose of the present work, one identifies $\psi$ with $\varphi$ and $-16\pi S$ with the right hand side of Eq.~(\ref{hhsclequ}) up to the quadratic order.
One should further truncate the series in the above expression at an appropriate order in the following discussion.

The leading contribution to $\varphi$ comes from the first term on the right hand side of Eq.~(\ref{hhsclequ}), which is the mass monopole moment,
\begin{equation}\label{vpmono1}
\begin{split}
  \varphi^{[1]}=&-\frac{1}{8\pi G_{4(0,0)}\zeta r}\int_\mathcal{M}\mathrm{d}^3x'T_*^{(1)}\\
  =&\frac{1}{8\pi G_{4(0,0)}\zeta r}\sum_am_aS_a.
  \end{split}
\end{equation}
It does not depend on time, so it does not contribute to the effective stress-energy tensor.

The next leading order term is the mass dipole moment,
\begin{equation}\label{vpdip1}
\begin{split}
  \varphi^{[1.5]}=&-\frac{1}{8\pi G_{4(0,0)}\zeta r}\partial_t\int_\mathcal{M}\mathrm{d}^3x'T_*^{(1)}\\
  =&\frac{1}{8\pi G_{4(0,0)}\zeta r}\partial_t\sum_am_aS_a(\hat n\cdot \vec x_a)\\
  =&\frac{1}{8\pi G_{4(0,0)}\zeta r}\sum_am_aS_a(\hat n\cdot\vec v_a).
  \end{split}
\end{equation}
This gives the leading contribution to the effective stress-energy tensor.

At the next next leading order, there are more contributions from the right hand side of Eq.~\eqref{hhsclequ}. First, there is the mass quadruple moment,
\begin{equation}\label{vpmquad1}
  \begin{split}
     \varphi_1^{[2]}= & -\frac{1}{8\pi G_{4(0,0)}\zeta r}\frac{\partial_t^2}{2}\int_\mathcal{M}\mathrm{d}^3x'T_*^{(1)}(\hat n\cdot\vec x_a)^2 \\
     = & \frac{1}{8\pi G_{4(0,0)}\zeta r}\frac{\partial_t^2}{2}\sum_am_aS_a(\hat n\cdot\vec x_a)^2 \\
       =&\frac{1}{8\pi G_{4(0,0)}\zeta r}\sum_am_aS_a[(\hat n\cdot\vec a_a)(\hat n\cdot\vec x_a)+(\hat n\cdot\vec v_a)^2].
  \end{split}
\end{equation}
The above three contributions \eqref{vpmono1}, \eqref{vpdip1} and \eqref{vpmquad1} all come from the first term in the source (the right hand side of Eq.~(\ref{hhsclequ})).

Other contributions to the scalar quadruple moment come from the remaining terms in the source. Firstly, there are the following three contributions,
\begin{eqnarray}
\varphi_2^{[2]}&=&-\frac{1}{8\pi G_{4(0,0)}\zeta r}\int_\mathcal{M}\mathrm{d}^3x'\left[G_{4(1,0)}T^{(2)}\right.\nonumber\\
&&\left.-2G_{4(0,0)}\left(\frac{\partial T}{\partial\phi}\right)^{(2)}\right]\nonumber\\
  &=&-\frac{1}{16\pi G_{4(0,0)}\zeta r}\sum_am_aS_av_a^2\nonumber\\
  &&+\frac{1}{64\pi^2G_{4(0,0)}^2\zeta r}\sideset{}{'}\sum_{a,b}\frac{m_am_b}{r_{ab}}\left(-\frac{3S_a}{2}\right.\nonumber\\
  &&\left.+\frac{3G_{4(1,0)}}{2G_{4(0,0)}\zeta}S_aS_b+\frac{S_a'S_b}{\phi_0\zeta}\right),\label{v22s}\\
\varphi_3^{[2]}&=&\frac{1}{16\pi G_{4(0,0)}\zeta}\int_\mathcal{M}\mathrm{d}^3x'\tilde hT_*^{(1)}\nonumber\\
&\approx&-\frac{1}{16\pi G_{4(0,0)}\zeta}\int_\mathcal{M}\mathrm{d}^3x'\tilde h_{00}T_*^{(1)}\nonumber\\
&=&\frac{1}{64\pi^2G_{4(0,0)}^2\zeta r}\sideset{}{'}\sum_{a,b}\frac{m_am_bS_b}{r_{ab}},\label{v23s}\\
\varphi^{[2]}_4&=&-\frac{1}{8\pi G_{4(0,0)}\zeta}\left(G_{4(2,0)}-\frac{G_{4(1,0)}^2}{G_{4(0,0)}}\right)\int_\mathcal{M}\mathrm{d}^3x'\varphi T^{(1)}\nonumber\\
  &=&\frac{1}{64\pi^2G_{4(0,0)}^2\zeta^2r}\left(G_{4(2,0)}-\frac{G_{4(1,0)}^2}{G_{4(0,0)}}\right)\times\nonumber\\
  &&\sideset{}{'}\sum_{a,b}\frac{m_am_bS_b}{r_{ab}},\label{v24s}
\end{eqnarray}
where $\sum'_{a,b}$ means summation over $a$ and $b$ with $a\ne b$, and in the second step of Eq.~(\ref{v23s}), the contribution from $\eta^{jk}\tilde h_{jk}T_*^{(1)}$ is dropped since it is of order $O(v^2)$ relative to $\tilde h_{00}T_*^{(1)}$.

Secondly, the term containing $T^{(1)}_{\mu\nu}\partial^\mu\partial^\nu\varphi$ in Eq.~(\ref{hhsclequ}) does not contribute as
\begin{equation}\label{t1ddp}
\begin{split}
  T^{(1)}_{\mu\nu}\partial^\mu\partial^\nu\varphi=&\stackrel{\rho O(1)\times O(v^2)}{T^{(1)}_{00}\partial^0\partial^0\varphi}\;+\;\stackrel{\rho O(v)\times O(v)}{2T^{(1)}_{0j}\partial^0\partial^j\varphi}\;\\
  &+\;\stackrel{\rho O(v^2)\times O(1)}{T^{(1)}_{jk}\partial^j\partial^k\varphi},
\end{split}
\end{equation}
where each term on the right hand side in the above expression indicates the relative order of that term to $T^{(1)}_{00}\varphi$, and $\rho=T^{(1)}_{00}$. Note that the action of $\partial^0$ increases the order by one since $\partial^0$ is actually $-\partial/c\partial t$.
Therefore, these terms are of higher order than those considered in Eqs. \eqref{v22s}, \eqref{v23s} and \eqref{v24s}, and will be ignored.
Similarly, the term containing $\tilde h_{\mu\nu}\partial^\mu\partial^\nu\varphi$ is also of higher order and dropped.

Thirdly, the following integral will be useful,
\begin{equation}\label{i1int}
  I_1 =\int_\mathcal{M}\mathrm{d}^3x\varphi T_*^{(1)}=-\frac{1}{8\pi G_{4(0,0)}\zeta}\sideset{}{'}\sum_{a,b}\frac{m_am_bS_aS_b}{r_{ab}}.
\end{equation}
The next useful integral is
\begin{equation}\label{i2int}
\begin{split}
   I_2 =& \int_\mathcal{M}\mathrm{d}^3x\varphi^2\\
   =&\frac{1}{64\pi^2 G_{4(0,0)}^2\zeta^2}\sum_{a, b}m_am_bS_aS_b\int_\mathcal{M}\frac{\mathrm{d}^3x}{r_ar_b}.
   \end{split}
\end{equation}
To compute it, we first consider the  terms with $a=b$,
\begin{equation}\label{i2int1}
\begin{split}
  I_{2,1}=&\frac{1}{64\pi^2 G_{4(0,0)}^2\zeta^2}\sum_{a}m_a^2S_a^2\int_{0}^\mathcal{R}\frac{\mathrm{d}^3x}{r_a^2}\\
  =&\frac{1}{16\pi G_{4(0,0)}^2\zeta^2}\sum_{a}m_a^2S_a^2\mathcal{R}.
  \end{split}
\end{equation}
Remember that $\mathcal{R}$ defines the boundary separating the near zone from the far zone. However, the scalar field should not depend on $\mathcal{R}$,
as shown in Ref.~\cite{Pati:2000vt}. So this result will be discarded. Second, consider the contributions from terms with $a\ne b$. Define $\vec y=\vec r_a=\vec x-\vec x_a$, then $\vec r_b=\vec x-\vec x_b=\vec y+\vec r_{ab}$.
Since the source is located deep inside the near zone, $|\vec x_a|\ll\mathcal{R}$. For $\vec x\in\mathcal{N}$, $|\vec x|^2=|\vec y+\vec x_a|^2=y^2+2\vec y\cdot\vec x_a+\vec x_a^2<\mathcal{R}^2$, and one knows that,
\begin{equation}\label{yrr}
  y<\mathcal{R}-\hat y\cdot\vec x_a+O(|\vec x_a|^2/\mathcal{R}),
\end{equation}
where $y=|\vec y|$ and $\hat y=\vec y/y$.
So
\begin{equation}\label{i2int2}
  \begin{split}
     \int_\mathcal{M}\frac{\mathrm{d}^3x}{r_ar_b} \approx &\int_\mathcal{M}\frac{\mathrm{d}^3y}{y|\vec y+\vec r_{ab}|}\\
     &-\oint_{\partial\mathcal{M}}\left.\frac{\vec x_a\cdot\hat y}{y|\vec y+\vec r_{ab}|}\right|_{y=\mathcal{R}}\mathcal{R}^2\mathrm{d}\Omega.
  \end{split}
\end{equation}
With the relation,
\begin{equation}\label{usef}
  \frac{1}{|\vec x-\vec x'|}=\sum_{l=0}^{\infty}\sum_{m=-1}^{l}\frac{4\pi}{2l+1}\frac{r_<^l}{r_>^{l+1}}Y^*_{lm}(\hat n)Y_{lm}(\hat n'),
\end{equation}
where $r_<$ is the smaller one of $r=|\vec x|$ and $r'=|\vec x'|$, $\hat n=\vec x/|\vec x|$ and $\hat n'=\vec x'/|\vec x'|$, one can show that the boundary integral should be
dropped as it depends on $\mathcal{R}$, and the first integral in Eq.~(\ref{i2int2}) gives $-2\pi r_{ab}$, independent of $\mathcal{R}$. Therefore,
\begin{equation}\label{i2fr}
  I_2=-\frac{1}{32\pi G_{4(0,0)}^2\zeta^2}\sideset{}{'}\sum_{a,b}m_am_bS_aS_br_{ab}.
\end{equation}

The third integral is,
\begin{equation}
\begin{split}
  I_3=&\int_\mathcal{M}(\partial_\mu\varphi)(\partial^\mu\varphi)\mathrm{d}^3x\\
  \approx&\int_\mathcal{M}(\partial_j\varphi)\partial^j\varphi\mathrm{d}^3x\\
  =&\oint_{\partial\mathcal{M}}\varphi\partial^j\varphi\mathrm{d} S_j-\int_\mathcal{M}\varphi\nabla^2\varphi\mathrm{d}^3x\\
  =&-\frac{1}{2G_{4(0,0)}\zeta}\int_\mathcal{M}\varphi T_*^{(1)}\mathrm{d}^3x,
\end{split}
\end{equation}
where $\mathrm{d} S_j$ is the surface area element. In the second step, $(\partial^0\varphi)\partial_0\varphi$ is ignored, as it is of higher order, and in the final step,
the boundary integral is discarded, as it depends on $\mathcal{R}$. The fourth integral is
\begin{equation}\label{i4int}
  \begin{split}
    I_4=&\int_\mathcal{M}(\partial_\mu\partial_\nu\varphi)(\partial^\mu\partial^\nu\varphi)\mathrm{d}^3x\\
    \approx&\int_\mathcal{M}(\partial_j\partial_k\varphi)(\partial^j\partial^k\varphi)\\
  =&\oint_{\partial\mathcal{M}}(\partial_k\varphi)\partial^j\partial^k\varphi \mathrm{d} S_j-\oint_{\partial\mathcal{M}}(\nabla^2\varphi)\partial^k\varphi\mathrm{d} S_k\\
  &+\int_\mathcal{M}(\nabla^2\varphi)^2\mathrm{d}^3x\\
  =&\int_\mathcal{M}\frac{(T_*^{(1)})^2}{4G_{4(0,0)}^2\zeta^2}\mathrm{d}^3x,
  \end{split}
\end{equation}
where in the second step, higher order terms $(\partial^0\partial^0\varphi)\partial_0\partial_0\varphi$ and $(\partial^0\partial^j\varphi)\partial_0\partial_j\varphi$ are ignored,
and in the final step, the surface integrals are discarded for the similar reasons as before. With this result, one can easily find out that the contribution of the second and the third line in Eq.~(\ref{hhsclequ}) vanishes.

Finally, the remaining contributions to the scalar field are
\begin{equation}
\begin{split}
  \varphi^{[2]}_{5}=& \frac{1}{4\pi r}\frac{G_{2(3,0)}}{2\zeta}\int_\mathcal{M}\mathrm{d}^3x'\varphi^2 \nonumber\\
  =&-\frac{1}{128\pi^2G_{4(0,0)}^2\zeta^2r}\frac{G_{2(3,0)}}{2\zeta}\sideset{}{'}\sum_{a,b}m_am_bS_aS_br_{ab},
  \end{split}
\end{equation}
and
\begin{equation}
\begin{split}
   \varphi_{6}^{[2]}=&-\frac{1}{4\pi r}\int_\mathcal{M}\mathrm{d}^3x'\Bigg\{\left[\frac{\varphi T_*^{(1)}}{G_{4(0,0)}\zeta}+(\partial_\mu\varphi)(\partial^\mu\varphi)\right]\\
  &\left(-\frac{G_{4(1,0)}}{2G_{4(0,0)}}+\frac{3G_{4(1,0)}^3}{2G_{4(0,0)}^2\zeta}-\frac{G_{2(1,1)}}{2\zeta}+\frac{G_{3(2,0)}}{\zeta}\right.\\
  &\left.-3\frac{G_{4(1,0)}G_{4(2,0)}}{G_{4(0,0)}\zeta}\right)+\frac{G_{4(1,0)}}{G_{4(0,0)}}(\partial_\mu\varphi)\partial^\mu\varphi\Bigg\}\\
   =&-\frac{1}{8\pi G_{4(0,0)} \zeta r}\Upsilon\int_\mathcal{M}\mathrm{d}^3x'\varphi T_*^{(1)}\\
   =&\frac{1}{64\pi^2G_{4(0,0)}^2\zeta^2 r}\Upsilon\sideset{}{'}\sum_{a,b}\frac{m_am_bS_aS_b}{r_{ab}},
   \end{split}
\end{equation}
where
$$\Upsilon=-\frac{3G_{4(1,0)}}{2G_{4(0,0)}}+\frac{3G_{4(1,0)}^3}{2G_{4(0,0)}^2\zeta}-\frac{G_{2(1,1)}}{2\zeta}+\frac{G_{3(2,0)}}{\zeta}-3\frac{G_{4(1,0)}G_{4(2,0)}}{G_{4(0,0)}\zeta}.$$
Add $\varphi_2^{[2]},\varphi_3^{[2]},\varphi_4^{[2]},\varphi_5^{[2]}$ and $\varphi_6^{[2]}$ together to give rise to Eq.~(\ref{varphi2}).


%

\end{document}